\documentclass[lettersize,journal]{IEEEtran}
\usepackage{amsmath,amsfonts}
\usepackage{amssymb}
\usepackage{algorithmic}
\usepackage{algorithm}

\usepackage{array}
\usepackage{textcomp}
\usepackage{stfloats}
\usepackage{url} 
\usepackage{verbatim}
\usepackage{graphicx}
\usepackage{cite}
\usepackage{bm}
\usepackage{microtype}
\usepackage{subcaption}
\usepackage{color}
\usepackage[table]{xcolor}
\usepackage[export]{adjustbox}
\newcommand{\be}{\begin{equation}}
\newcommand{\ee}{\end{equation}}
\newcommand{\ba}{\begin{array}}
\newcommand{\ea}{\end{array}}
\newcommand{\bea}{\begin{eqnarray}}
\newcommand{\eea}{\end{eqnarray}}

\newcommand{\diag}{{\operatorname{diag}}}
\newcommand{\herm}{^{\mbox{\scriptsize \emph{H}}}}  
\newcommand{\tran}{^{\mbox{\scriptsize \emph{T}}}}  

\newcommand{\tensor}[1]{\bm{\mathcal{#1}}}

\DeclareCaptionFont{blue}{\color{blue}}
\DeclareCaptionFont{black}{\color{black}}
\begin{document}
\title{Efficient Channel Prediction for Beyond Diagonal RIS-Assisted MIMO Systems with Channel Aging

\author{Nipuni  Ginige, \IEEEmembership{Graduate Student Member, IEEE}, Arthur Sousa de Sena, \IEEEmembership{Member, IEEE}, Nurul Huda Mahmood, Nandana Rajatheva, \IEEEmembership{Senior Member, IEEE} and
		 Matti Latva-aho, \IEEEmembership{Fellow, IEEE} }
\thanks{Partial results of this work was presented at the 2024 IEEE Wireless Communications and Networking Conference (WCNC) Workshops\cite{Nip_CNN_AR}. This research was supported by the Research Council of Finland (former Academy of Finland) 6G Flagship Programme (Grant Number: 346208).}
\thanks{Nipuni Ginige, Arthur S. de Sena, Nurul H. Mahmood, Nandana Rajateva, and Matti Latva-aho are with the Centre for Wireless Communications, University of Oulu, 90570 Oulu, Finland (e-mail: nipuni.ginige@oulu.fi, arthur.sena@oulu.fi, nurulhuda.mahmood@oulu.fi, nandana.rajatheva@oulu.fi, matti.latva-aho@oulu.fi).}
}

\maketitle

\begin{abstract}
Novel reconfigurable intelligent surface (RIS) architectures, known as beyond diagonal RISs (BD-RISs), have been proposed to enhance reflection efficiency and expand RIS capabilities. However, their passive nature, non-diagonal reflection matrix, and the large number of coupled reflecting elements complicate the channel state information (CSI) estimation process. The challenge further escalates in scenarios with fast-varying channels.
In this paper, we address this challenge by proposing novel joint channel estimation and prediction strategies with low overhead and high accuracy for two different RIS architectures in a BD-RIS-assisted multiple-input multiple-output system under correlated fast-fading environments with channel aging. The channel estimation procedure utilizes the Tucker2 decomposition with bilinear alternative least squares, which is exploited to decompose the cascade channels of the BD-RIS-assisted system into effective channels of reduced dimension. The channel
prediction framework is based on a convolutional neural network combined with an autoregressive predictor. The estimated/predicted CSI is then utilized to optimize the RIS phase shifts aiming at the maximization of the downlink sum rate. Insightful simulation results demonstrate that our proposed approach is robust to channel aging, and exhibits a high estimation accuracy. Moreover, our scheme can deliver a high average downlink sum rate, outperforming other state-of-the-art channel estimation methods. The results also reveal a remarkable reduction in pilot overhead of up to 98\% compared to baseline schemes, all imposing low computational complexity.

\end{abstract}


\section{Introduction}

\IEEEPARstart{R}{econfigurable} intelligent surface (RIS) has become a promising technology to enhance the performance of sixth-generation (6G) and beyond systems, offering the potential to manipulate the electromagnetic propagation environment through the configuration of nearly passive reflecting elements. An RIS comprises a large number of low-cost elements that are capable of inducing phase changes to incoming signals toward diverse objectives, including improving the received signal strength in scenarios where receivers are experiencing weak signal reception or when the direct path between the transmitter and receiver is completely blocked. RISs function as a smart reflection hub that can support massive connectivity, manage interference through strategic passive beamforming, and mitigate security vulnerabilities\cite{Liu21_RIS_survey}.

In a conventional RIS, each reflecting element is connected to a distinct reconfigurable self-impedance. This allows the elements to operate independently from each other, which results in a diagonal reflection matrix. Conventional RIS has low flexibility in wave manipulation since it can only manipulate the diagonal entries of the reflection matrix. To improve the reflection efficiency and overall performance, beyond diagonal RIS (BD-RIS) was introduced \cite{BD_RIS_Intro}. The key feature of a BD-RIS is the incorporation of inter-element connections, which, although it introduces additional circuit complexity, enables a more flexible and powerful signal optimization framework. BD-RISs provide several benefits over conventional RISs, including enhanced wave manipulation and superior reflection performance with lower numbers of reflecting elements\cite{BD_RIS_Intro}. Two main BD-RIS architectures have been proposed, namely fully-connected RIS and group-connected RIS. The reflecting elements in a fully-connected RIS are interconnected with every other element, while the reflecting elements in a group-connected RIS are interconnected within each group but independent across different groups. Fully-connected RISs offer the highest performance at the cost of the highest circuit complexity. Group-connected RISs, on the other hand, exhibit an intermediate performance but count with moderate complexity, providing a bridge between conventional and fully-connected RISs \cite{BD_RIS_archi,BD_RIS_arch,de2024beyond}.

Channel estimation plays a crucial role in modern multiple-input multiple-output (MIMO) systems, being the foundation for various signal processing and precoding techniques essential to enable features such as spatial diversity and multiplexing, interference mitigation, and dynamic adaptation to fast-varying wireless environments. The accuracy of the channel state information (CSI) has, consequently, a direct impact on the system performance, spectral efficiency, and reliability. The importance of channel estimation to BD-RIS-assisted MIMO systems is even more pivotal as, in addition to traditional active base station (BS) precoders, the passive RIS reflecting coefficients need to be carefully optimized based on the additional RIS reflected channels, which are generally challenging to acquire. Channel estimation for BD-RIS-assisted MIMO systems, for both the fully-connected and group-connected architectures, is practically more difficult since the number of parameters that need to be estimated can be very large due to the non-diagonality of the associated reflection matrix \cite{BD_RIS_Intro}. This challenge further escalates in time-varying environments, in which channel coefficients varying according to Doppler shift changes due to user movements, a phenomenon known as channel aging \cite{YuanChannelAginML}.

Generally, channel models have a multidimensional algebraic structure which causes challenges in channel estimation by rendering it a large-scale estimation problem. Multilinear algebra and tensor analysis offer advanced tools to tackle the multidimensional nature of CSI acquisition-related problems. A tensor is a multidimensional array. Higher-order tensor decompositions and applications were discussed by the authors in \cite{TensorReview}. Tensor decompositions are applied to extract and explain the properties of data arrays \cite{TensorReview}. The PARAFAC and Tucker decomposition are two main tensor decomposition approaches, with applications in many different fields, including wireless communication\cite{ Tensor_wireless_communication, TensorbasedCERelay, TensorforRelay, DEALMEIDA20092103}. Tensor decomposition techniques have been exploited as efficient methods for estimating large channel matrices due to their favorable features, such as their uniqueness and their ability to offer low overhead and low-complexity estimation of low-rank matrices \cite{A_PARAFAC_RIS_CE, TensorbasedCERelay}. 

\subsection{Related Works}
Tensor decomposition was used for channel estimation in two-hop vehicle-to-everything MIMO in \cite{Tucker2_V2X} and for joint channel estimation and signal detection in MIMO relay systems in \cite{TensorbasedCERelay, TensorforRelay}. The PARAFAC decomposition was used for channel estimation in conventional RIS multi-user multi-input-single-output system in \cite{PARAFAC_CE_RIS, PARAFAC_RIS_CE2,A_PARAFAC_RIS_CE}. {Furthermore, the authors in \cite{wang2024tensor} used PARAFAC decomposition for channel estimation in  MIMO orthogonal frequency division multiplexing (OFDM) systems in high-mobility scenarios.}  Moreover, alternating least squares with tensor decomposition methods such as Tucker2 and nested PARAFAC were used for channel estimation in conventional double-RIS-assisted MIMO systems in \cite{Tucker2_double_RIS, nested_PARAFAC_double_RIS}. The authors in \cite{Tensor_based_CE_RIS} exploited the multi-linear structure of the nested PARAFAC model to jointly estimate the time-varying phase noise and channels in a conventional RIS-assisted system.  Two tensor-based algorithms were proposed for joint channel and RIS imperfections estimation in a conventional RIS-assisted MIMO system by the authors in \cite{Tensor_imperfection_CE_RIS}. A three-stage joint channel decomposition and prediction framework using deep neural networks (DNN) was proposed for conventional RIS-assisted systems by the authors in \cite{CE_RIS_decomposition_prediction_DNN}.

 The authors of \cite{JiangChannelAgingImpactRIS} theoretically analyzed the impact of channel aging on conventional RIS-assisted systems. An investigation of the impact of channel aging on the performance of conventional RIS-assisted massive MIMO systems under both spatial correlation and imperfect CSI conditions was presented in \cite{AnastasiosChannelAgingRIS}. The authors in \cite{ZhangChannelAgingPrecodingRIS} proposed an efficient sub-optimal channel aging-aware precoding algorithm for conventional RIS-aided multi-user communications.  All existing works show that channel aging can cause significant performance degradation in communication systems.  Moreover, channel aging can further increase the pilot overhead since more frequent estimations need to be carried out. The authors in \cite{Hu2TimeScaleCE_RIS} proposed a two-timescale estimation framework for conventional RIS-assisted systems, capable of reducing the pilot overhead by assuming that the channel between the BS and the RIS is quasi-static and user-associated channels are time-varying.
{Furthermore, the authors in \cite{RIS_CE_mmwave} exploited the sparsity and the correlation of multiuser cascaded channels in millimeter-wave MISO systems for a low pilot overheat channel estimation strategy. }A channel decomposition and prediction framework based on the long short-term memory (LSTM) neural network (NN) architecture was proposed by the authors in \cite{CE_RIS_decomposition_prediction_DNN}. Both works assumed that the system operates under full-duplex mode. The authors of \cite{AR_model} examined the autoregressive (AR) modeling approach for the accurate prediction of correlated Rayleigh time-varying channels. A machine learning (ML)-based framework was proposed to improve the CSI prediction quality in \cite{YuanChannelAginML}. The authors of this work implemented a convolutional neural network (CNN) to identify the channel aging pattern and combined it with the AR method to predict the CSI in MIMO systems. The work in \cite{Nip_CNN_AR} proposed a channel prediction framework for conventional RIS-assisted MIMO systems based on a CNN integrated with an AR predictor. The least squares (LS)-based estimator that achieves the lower bound of the mean square error (MSE) for group-connected RISs, at the cost of an overwhelming pilot overhead, was proposed by the authors in \cite{li2023channel}.  {Further, they propose an efficient pilot sequence and BD-RIS design to achieve the minimum MSE in \cite{BD-RIS-CE2}.} 


\subsection{Motivation and Contributions}

Optimizing the RIS reflection matrix requires accurate CSI. Adopting conventional CSI estimation methods incurs significant overhead and computational complexity, which is further exacerbated when considering BD-RIS-assisted MIMO systems. Low-overhead and efficient channel estimation schemes for BD-RIS-assisted MIMO systems, considering both fully-connected and group-connected architectures, are yet to be developed. Moreover, individual knowledge of user equipment (UE)-RIS and RIS-BS channels leads to low complexity RIS reflection matrix optimization. However, conventional channel estimation methods only estimate the composite channel. {Tensor decomposition-based channel estimation shows good performance in conventional RIS-assisted systems.} Even though PARAFAC decomposition was used to decompose and estimate UE-RIS and RIS-BS channels in conventional RIS-assisted systems {with excellent performance} \cite{PARAFAC_CE_RIS, PARAFAC_RIS_CE2,A_PARAFAC_RIS_CE}, it is not applicable for BD-RIS-assisted systems due to the non-diagonal nature of the reflection matrix. {However, tensor-based channel estimation for BD-RIS has not yet been explored.} The CNN-AR channel estimation method introduced in \cite{Nip_CNN_AR} exhibited promising performance in terms of accuracy and pilot overhead for conventional RIS-assisted MIMO systems. Nevertheless,  to the best of our knowledge, the generalization of this method to BD-RISs remains unexplored. These major literature gaps motivate the development of this work.

In this paper, we propose joint channel estimation and channel prediction methods for fully-connected and group-connected BD-RIS-assisted MIMO systems with channel aging. We show that the received signal in BD-RIS assisted MIMO systems can be arranged into a third-order tensor which follows Tucker2 decomposition \cite{TensorReview}. {Moreover, alternative least-squares is an efficient method to compute Tucker decompositions with low training overhead and high accuracy \cite{TensorReview}. }   Hence, we propose a bilinear alternative least squares (BALS) based scheme with Tucker2 decomposition for channel estimation of BD-RIS-assisted systems. Moreover,  we propose to integrate it with CNN-AR-based channel prediction. The main contributions of this work are further elaborated in the following.
\begin{itemize}

    \item Harnessing the advantages of RIS in improving the received signal strength requires optimizing the RIS reflection matrix by aligning them to the transmitter-RIS and RIS-receiver channels. This requires accurate CSI estimation, which is rather challenging due to the passive nature of the RIS elements and the high dimensionality of the data~\cite{Swindlehurst22_RISchannelEstimation}. Some authors have proposed to circumvent the need for CSI estimation by applying deep reinforcement learning-based techniques to directly learn the optimal reflection matrix~\cite{hashemi23_DRL_RIS}. However, this leads to slow convergence and a high training overhead. We propose to overcome this limitation by proposing a deep unfolding approach~\cite{Stimming19_deepUnfolding}, which combines the optimization algorithms with ML tools to efficiently solve complex problems that cannot be readily solved using conventional methods. More precisely, we propose an optimized pilot-based channel estimation method and a CNN-based channel prediction method to accurately acquire the CSI, which is then used as inputs in an alternating optimization algorithm to optimize the RIS reflection matrix with the objective of maximizing the downlink sum rate.
    
    \item We propose a novel BALS-based channel estimation scheme with Tucker2 decomposition (Tucker2-BALS) for BD-RIS-assisted MIMO systems. Our proposed Tucker2-BALS algorithm can estimate the channels in a BD-RIS-assisted system with significantly low pilot overhead and high accuracy compared to conventional LS. The proposed Tucker2-BALS channel estimation algorithm can be applied to both the fully-connected and group-connected architectures. Moreover, the superiority of our proposed Tucker2-BALS algorithm increases with the capability of estimating the BS-RIS and RIS-UE channels separately rather than estimating the composite channel since it reduces the number of elements that need to be estimated and the knowledge of individual channels leads to low complexity channel estimation and prediction, and reflection matrix optimization. {Unlike the PARAFAC decomposition which is unique under permutation and scaling ambiguities conditions, Tucker decompositions are not essentially unique since unknown non-singular matrices can act as loading matrices\cite{TensorReview, TensorbasedCERelay}. Therefore, we present the uniqueness and identifiability conditions for the Tucker2 decomposition. }

    \item   The proposed Tucker2-BALS estimation strategy is integrated into a channel prediction framework to address channel aging and further reduce pilot overhead by skipping frequent pilot training. Specifically, a CNN model is designed and trained to identify the aging characteristics of the correlated time-varying wireless channels in a multi-user scenario. The obtained aging pattern is then integrated with the AR approach to forecast the desired CSI.
    
    \item Insightful simulation results are provided to assess the performance of the proposed model, revealing remarkable estimation and prediction accuracy. Furthermore, we have shown that a higher downlink sum rate can be achieved using the estimated CSI by our proposed channel estimation scheme than the conventional method. Moreover, our proposed joint channel estimation and channel prediction algorithm for BD-RIS-assisted MIMO systems achieve significantly lower pilot overhead and low computational complexity compared to the considered baseline channel estimation method for BD-RIS.  
\end{itemize}

The rest of the paper is organized as follows. The system model for fully-connected and group-connected BD-RIS is described in Section \ref{SM}. In Sections \ref{channel estimation} and \ref{channel prediction}, we present the proposed Tucker2 decomposition-based channel estimation algorithm and CNN-based channel prediction model, respectively. The performance metrics used to analyze our proposed schemes are described in Section \ref{performance}. The numerical results are presented in Section \ref{results}, and Section \ref{conclusion} concludes our paper.

\subsubsection*{Notations and Properties}
\label{notations}
We use lower-case letters for scalars, bold-faced lower-case letters for column vectors, bold-faced upper-case letters for matrices, and bold-faced upper-case calligraphic letters for tensors of order higher than two. Notations for transpose, Hermitian transpose, Moore-Penrose pseudo-inverse, and Frobenius norm of $\mathbf{A}$ are $\mathbf{A}\tran, \mathbf{A}\herm $, $\mathbf{A}^\dagger$, and $||\mathbf{A}||_F$, respectively. $[\mathbf{A}]_{i,j}$ denotes the $(i,j)$-th element of $\mathbf{A}$, while $[\mathbf{A}]_{i,:}$ and $[\mathbf{A}]_{:,j}$ denote $i$-th row and $j$-th column of $\mathbf{A}$, respectively. The operator $\operatorname{vec} (\cdot)$ vectorizes its input matrix while the operators $\otimes$ and $\odot$ define the Kronecker matrix product and Hadamard product, respectively. The operator $\diag(\cdot)$ transforms a vector into a diagonal matrix and $\operatorname{bdiag}(\cdot)$ structures input matrices into a block-diagonal matrix. The operator $\operatorname{circshift}(\mathbf{a}, N)$ circularly shifts vector $\mathbf{a}$ by $N$ positions. $\mathbf{I}_N$ is the $N\times N$ identity matrix and $\mathbf{1}_{N,M}$ is the $N \times M$ matrix of ones.  $[\tensor{A}]_{(n)}$ represents the mode-$n$ unfolding (matricization) of $\tensor{A}$ along its $n$-th mode. Multiplication between a tensor and a matrix in mode $n$ is defined as the $n$-mode product. The $n$-mode (matrix) product of $\tensor{A}$ with $\mathbf{B}$ is denoted by $\tensor{C} = \tensor{A} \times_n \mathbf{B}$ and its mode-$n$ unfolding is $[\tensor{C}]_{(n)}= \mathbf{B}[\tensor{A}]_{(n)}$. In this paper, we use the following Kronecker product property:
\begin{equation}\operatorname{vec} \left(\mathbf{ABC}^{\tran} \right) = \left(\mathbf{C} \otimes \mathbf{A} \right) \operatorname{vec} \,(\mathbf{B}) \label{kronerkerI}.\end{equation}

\section{System Model}
\label{SM}

\begin{figure}[t]  
    \centering
    \includegraphics[trim={1cm 0 -1cm 0},width=0.5\textwidth]{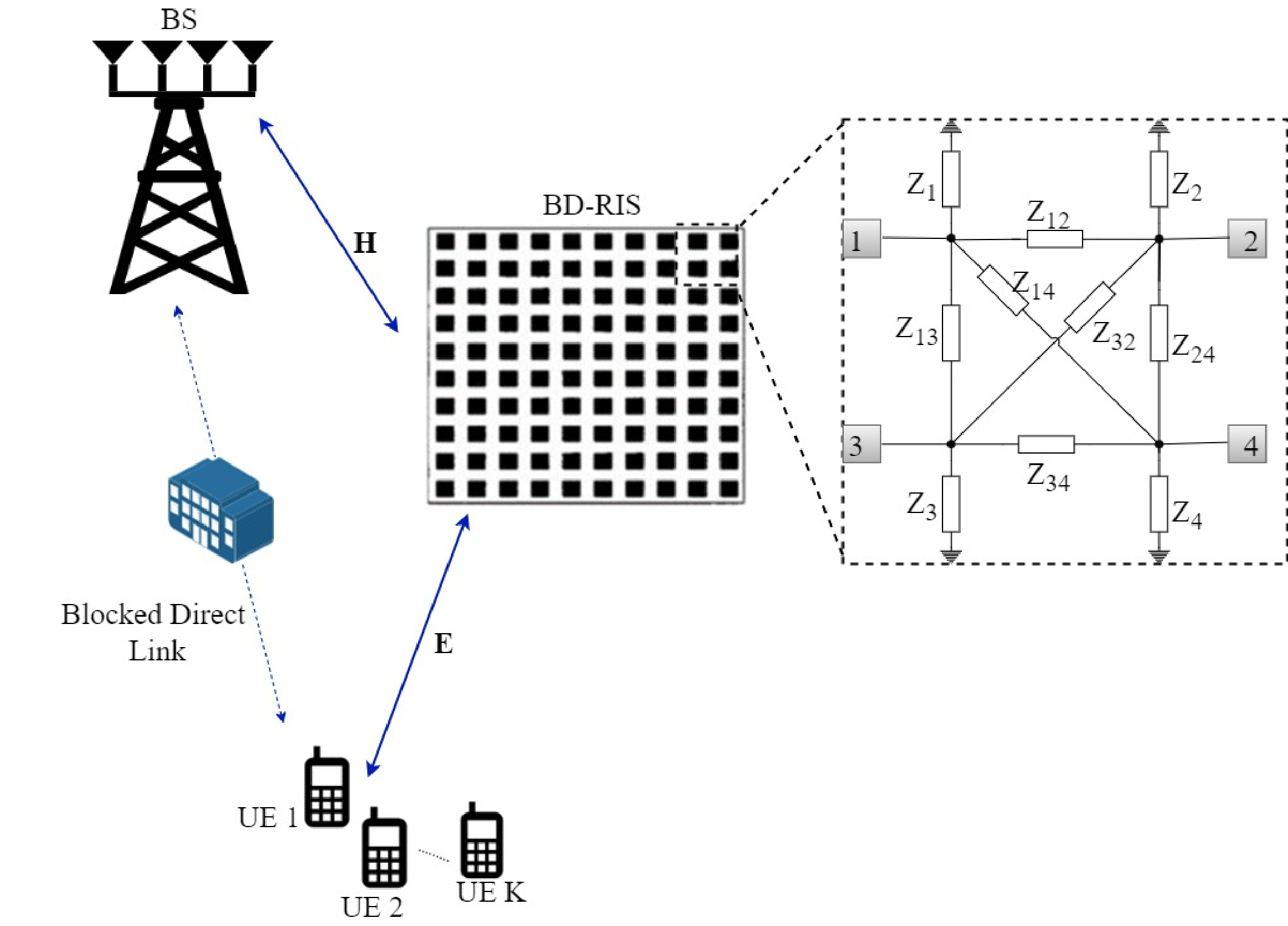}
    \caption{System model. A BD-RIS is deployed to assist transmissions between a multi-antenna BS and multiple users. The figure shows a section of the BD-RIS with four fully connected elements. Direct paths between UEs and the BS are not available.}
    \label{fig:Illustration of the system model}
  
\end{figure} 

We consider a BD-RIS-assisted MIMO system, as shown in Fig.~\ref{fig:Illustration of the system model}, where a BS equipped with $N$ transmit antennas communicates with $K$ single-antenna UEs.  A BD-RIS composed of  $M $  reflecting elements is positioned between the BS and UEs. We assume that the direct paths between UEs and BS are not available.

The BD-RIS is equipped with an intelligent controller that dynamically fine-tunes the reflection coefficients of the RIS elements based on the instantaneous CSI obtained through periodic estimation. Specifically, we denote by $\mathbf{E}[l] =\bigl[\mathbf{e}_1[l], \ldots \mathbf{e}_K[l]\bigr] \in \mathbb{C}^{M\times K}$ the RIS-UE channel matrix in $l$-th coherence interval of duration $T_c$, where $\mathbf{e}_k[l]\in \mathbb{C}^{M\times 1}$, for $k = 1, \cdots, K$, is the RIS-UE channel vector for the $k$-th UE in the $l$-th coherence interval. {Here $T_c = KT+T_d+T_u$, where $T$ is the optimal number of {training blocks} needed in which each block has $\tau \geq K$ time slots. Here we consider $\tau =K$. }Furthermore, $T_d$ and $T_u$ are time slots for downlink and uplink data transmission, respectively. Moreover, we represent by $\mathbf{H}\in {\mathbb{C}^{N \times M}}$  the RIS-BS channel matrix. In this work, we assume that channel aging does not affect the RIS-BS channels, which is motivated by the fact that the BS and the RIS are usually fixed, making the associated channels quasi-static. Due to this reason, channel aging is modeled only in the UE-RIS channels.

 While previous research, e.g., \cite{li2023channel}, has operated under the assumption of an independent Rayleigh fading model, real-world scenarios often exhibit spatially and temporally correlated fading, which affects the performance \cite{AnastasiosChannelAgingRIS}. We take this important property under consideration and model $\mathbf{e}_k[l]$ with correlated entries, as follows
\begin{equation}\mathbf{e}_k [l] = {\sqrt{\beta_e}}{\mathbf {R}}_{\text{RIS}}^{1/2}{\mathbf {q}}_{{e,k}}e^{j2\pi f_{\mathrm{d}}T_sl},\end{equation}
where ${\mathbf {R}}_{\text{RIS}} \in {\mathbb{C}^{M \times M}}$ denote the deterministic Hermitian-symmetric positive semi-definite correlation matrix at the RIS with $\text{tr}({\mathbf {R}}_{\text{RIS}})=M$,  ${\mathbf {q}}_{{e,k} } $ represents the corresponding  {Rician} fading vector and  {$\sqrt{\beta_e}$ is the large-scale fading (which includes path loss and shadowing) corresponds to the RIS-UE channel.} Moreover, $T_s$ is the sampling duration and  $f_{\mathrm{d}}$ is the maximum Doppler frequency which is given by $f_{\mathrm{d}}=vf_{c}/c$, in which $v$ is the velocity of the $k$-th UE, $c$ is the speed of light, and $f_{c}$ is the carrier frequency. It is important to note that the correlation matrix $\mathbf{R}_{\text{RIS}}$ remains constant over a larger number of coherence intervals. 

The correlated  {Rician} RIS-BS channel matrix $\mathbf{H}$  is modeled as   \begin{align}
  \mathbf{H} =&  {\sqrt{\beta_H}}{\mathbf {R}}_{\text{BS}}^{1/2}{\mathbf {Q}}_{{H} }{\mathbf {R}}_{\text{RIS}}^{1/2},
\end{align}
where ${\mathbf {R}}_{\text{BS}} \in {\mathbb{C}^{N \times N}} $ denote the deterministic Hermitian-symmetric positive semi-definite correlation matrix of the BS with  $\text{tr}({\mathbf {R}}_{\text{BS}})=N$, ${\mathbf {Q}}_{{H} } $ is the {Rician} fading matrix which remains constant for a larger number of coherence intervals and  {$\sqrt{\beta_H}$  is the large scale fading corresponds to the RIS-BS channel.} 
 
\begin{figure}[t]  
	\centering
	\includegraphics[trim={-1cm 2cm 4cm 1cm},width=0.5\textwidth]{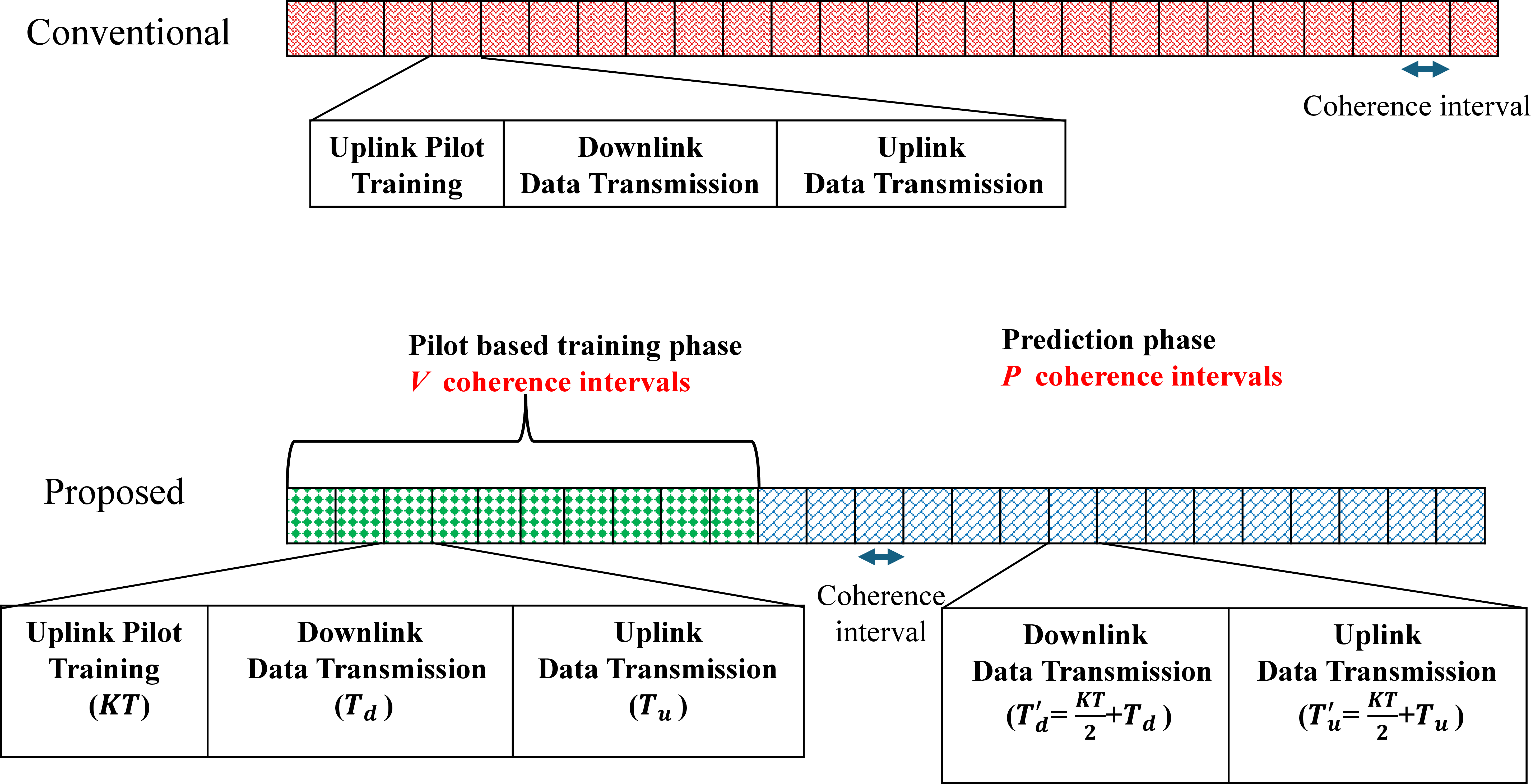 }
	\caption{{Simplified transmission block: Conventional versus proposed channel estimation scheme. In the pilot-based training phase, Tucker2-BALS is used to acquire CSI. In the prediction phase, uplink pilot training is removed due to the introduction of ML-based CSI prediction.}}
	\label{fig:Transmission Block}
 \end{figure}

We adopt a time-division duplexing (TDD) protocol, which enables the use of the same frequency for both uplink and downlink transmissions, in which we leverage channel reciprocity for acquiring the initial CSI and subsequently predicting it. In our proposed channel estimation scheme, the transmission block is composed of a pilot-based training phase and a prediction phase, as shown in Fig.~\ref{fig:Transmission Block}.

\begin{figure}[t]  
	\centering
	\includegraphics[trim={0cm 0cm -1cm 0cm},width=0.5\textwidth]{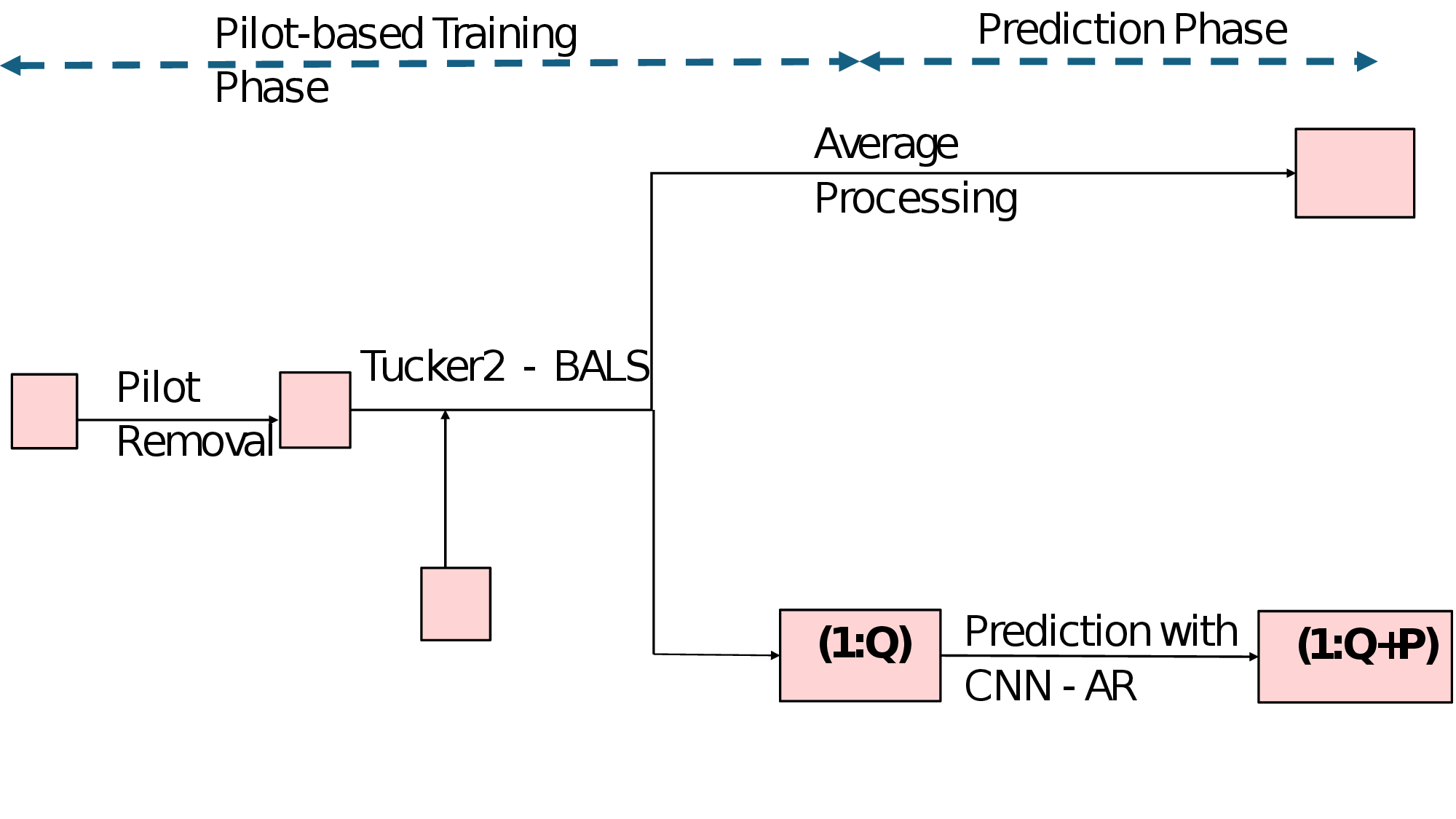 }
	\caption{System architecture: Proposed joint channel estimation and channel prediction scheme for BD-RIS.}
	\label{fig:sys arch}
 \end{figure}

In conventional TDD-based approaches, within each coherence interval, pilot symbols are sent in the uplink to estimate the reflected RIS channels. The estimated CSI is used for downlink and uplink data transmission by assuming that the channel is static during one coherence interval and that channel reciprocity is satisfied. In the proposed channel estimation scheme, the pilot-based training phase, consisting of $V$ coherence intervals, works similarly as in conventional TDD. During the pilot-based phase, we adopt a Tucker2-BALS algorithm to estimate the reflected channels to and from the RIS instead of LS estimation. During the prediction phase, on the other hand, the channel estimates acquired in the first phase are exploited to predict the CSI through ML in the following $P$ coherence intervals. The predicted CSI is then used for optimize the RIS reflection matrix and active beamforming matrix with the objective of maximizing the sum rate for both downlink and uplink data transmission. Fig.~\ref{fig:sys arch} shows the system architecture of the proposed channel estimation scheme.  

Conventional RIS  with $M$ reflecting elements has a diagonal reflection matrix, i.e., $\bm{\Theta} = \diag{(e^{j\theta_1},\cdots,e^{j\theta_M})}$. In contrast, BD-RIS has a full reflection matrix which is categorized as fully-connected and group-connected according to the structure of the RIS reflection matrix.

\subsection{Fully-connected RIS}
A fully-connected RIS with $M$ reflecting elements can be considered an $M$-port reconfigurable impedance network where every reflecting element is interconnected with every other reflecting element \cite{PassiveBeamforming_BD_RIS}. The reflection matrix of a fully-connected RIS in the $t$-th {training block} can be expressed as $\bm{\Theta}_t \in \mathbb{C}^{M \times M}$, in which the constraints $ \bm{\Theta}_t = \bm{\Theta}_t\tran$ and $ \bm{\Theta}_t\bm{\Theta}\herm_t= \mathbf{I}_M$ need to be satisfied. The UE-RIS-BS cascade channel in the $l$-th coherence interval for the fully-connected RIS can be expressed as
\begin{equation}
\mathbf{G}_t[l] = \mathbf{H}\bm{\Theta}_t\mathbf{E}[l].
\label{FullyCas}
\end{equation} 

We consider that the channel remains constant throughout each coherence interval but experiences variations from one interval to the next.  Moreover, the channels within a specific coherence interval are correlated with the channels in preceding intervals. 

\subsection{Group-connected RIS}

 A group-connected RIS comprising $M$ reflecting elements can be treated as an $M$-port group-connected reconfigurable impedance network where every reflecting element is interconnected with every other reflecting element within its own group but independent across other groups. We assume that there are $G$ groups, each containing $\bar{M}=M/\bar{G}$ reflecting elements.  The reflection matrix for a group-connected RIS in the $t$-th {training block} can be expressed as $\bm{\Theta}_t = \operatorname{bdiag}(\bm{\Theta}_{t,1}, \ldots, \bm{\Theta}_{t,\bar{G}}) \in \mathbb{C}^{M \times M}$, where $\bm{\Theta}_{t,g} \in \mathbb{C}^{\bar{M}\times\bar{M}}$ is the reflection matrix for the $g$-th group, for $g=1,\ldots,\bar{G}$, in the $t$-th {training block}, in which we should obey the following constraints $\bm{\Theta}_{t,g} = \bm{\Theta}_{t,g}\tran$ and $ \bm{\Theta}_{t,g}\bm{\Theta}_{t,g}\herm= \mathbf{I}_{\bar{M}}$. The UE-RIS-BS cascade channel in the $l$-th coherence interval for the group-connected RIS can be expressed as
\begin{align}
\mathbf{G}_t[l] = \mathbf{H}\bm{\Theta}_t\mathbf{E}[l] &= \sum_{g \in \bar{G}} \mathbf{H}_g\bm{\Theta}_{t,g}\mathbf{E}_g[l], \label{groupcas}
\end{align} 
 where $\mathbf{H}_g=[\mathbf{H}]_{:,(g-1)\bar{M}+1:g\bar{M}}$, and $\mathbf{E}_g[l]=[\mathbf{E}[l]]_{(g-1)\bar{M}+1:g\bar{M},:} $, for $ g = 1,\ldots,\bar{G}$.

\section{Channel Estimation for BD-RIS}
\label{channel estimation}
In this Section, we focus on the pilot-based training phase illustrated in Fig.~\ref{fig:Transmission Block}. Predefined reflection matrices are used in the BD-RIS during channel estimation. Moreover, orthogonal pilot sequences are sent by the users during the pilot-based training phase. {The transmitted signals during the $t$-th training block are represented by $\mathbf{X}_t \in \mathbb{C}^{K \times K} $ for $t = 1, \cdots, T$, with $KT< T_c$, where $\mathbf{X}_t\mathbf{X}_t\herm=\mathbf{I}_K$.}

The received signal  in the $t$-th {training block} in the $l$-th coherence interval is
\begin{equation} {\mathbf{Y}}_t[l]= {\sqrt{P_p}}\left( {\mathbf{H} \bm{\Theta}_t \mathbf{E}[l] } \right){\mathbf{X}}_t  + \mathbf{V}_t[l],\end{equation}
where ${P_p}$ is the power used to transmit pilots by the UEs and  $\mathbf{{V}}_t[l] \sim {\mathcal{CN}}\left( {0,{\sigma_n^2}{\mathbf{I}_{N\times K}}} \right)$ is the complex white Gaussian noise matrix. The full UE-RIS-BS cascade channel matrix, including the reflection RIS matrix, can be then estimated as follows
\be \mathbf{\tilde{G}}_t[l] = ({\sqrt{P_p}})^{-1} {\mathbf{Y}}_t[l]{\mathbf{X}}_t\herm =\underbrace {\mathbf{H} \bm{\Theta}_t \mathbf{E}[l] }_{\triangleq{\mathbf{G}_t[l]}} + \mathbf{\tilde{V}}_t[l], \label{noisycas}
\ee
where $\mathbf{\tilde{G}}_t[l]$ is the noisy version of the ${\mathbf{G}_t[l]}$ and $ \mathbf{\tilde{V}}_t[l] \triangleq ({\sqrt{P_p}})^{-1} {\mathbf{V}}_t[l]{\mathbf{X}}_t\herm  \in \mathbb{C}^{N \times K}$ is the noise matrix after removing the pilot symbols.

\subsection{Conventional LS-Based Channel Estimation}

Before introducing our proposed tensor-based strategies, we provide in this subsection details of conventional LS channel estimation schemes for both fully-connected and group-connected RISs, which are used as baselines in our simulation results.

By using the Kronecker identity in \eqref{kronerkerI}, we can transform \eqref{FullyCas} and \eqref{groupcas} into \eqref{cascadeCH} and \eqref{groupcas2}, respectively, as follows
\be \operatorname{vec}(\mathbf{G}_t[l])= \underbrace{(\mathbf{E}[l]\tran\otimes\mathbf{H})}_{=\mathbf{Z}[l]\in \mathbb{C}^{NK \times M^2}}\operatorname{vec}(\bm{\Theta}_t),\label{cascadeCH} \ee where we have defined $\mathbf{Z}[l] = \mathbf{E}[l]\tran\otimes \mathbf{H}$, and \be \operatorname{vec}(\mathbf{G}_t[l])=\sum_{g \in \bar{G}}\underbrace{(\mathbf{E}_g[l]\tran\otimes\mathbf{H}_g)}_{=\mathbf{Z}_g[l] \in \mathbb{C}^{NK \times \bar{M}^2}}\operatorname{vec}(\bm{\Theta}_{t,g}), \label{groupcas2} \ee
 {where we defined $\mathbf{Z}_g[l] = \mathbf{E}_g[l]\tran \otimes \mathbf{H}_g$ and $\tilde{\mathbf{Z}}[l] = \bigl[\mathbf{Z}_1[l], \ldots, \mathbf{Z}_{\bar{G}}[l] \bigr] \in \mathbb{C}^{NK \times \bar{G}\bar{M}^2}$.}

Conventional LS-based channel estimation algorithm for group-connected RIS proposed by the authors in \cite{li2023channel} can be described as follows: Once the estimate of the full channel, $\mathbf{\tilde{G}}_t[l]$, in \eqref{noisycas} is obtained, the LS estimation of the channel matrix $\mathbf{\tilde{Z}}[l]$ defined in  \eqref{groupcas2}  can be computed by
\be \mathbf{\hat{Z}}[l]= \mathbf{\Upsilon}[l]\bm{\Phi}^{\dagger} = \mathbf{\Upsilon}[l]\bm{\Phi}\herm(\bm{\Phi}\bm{\Phi}\herm)^{-1}, \ee
where $\mathbf{\Upsilon}[l]=\bigl[\operatorname{vec}(\mathbf{\tilde{G}}_1[l]),\ldots,\operatorname{vec}(\mathbf{\tilde{G}}_{T}[l]) \bigr] \in \mathbb{C}^{NK \times T}$ and $\bm{\Phi}= [\bm{\phi}_1, \ldots, \bm{\phi}_{T}] \in \mathbb{C}^{\bar{G}\bar{M}^2\times T},$ where $\bm{\phi}_t =  \bigl[ [\operatorname{vec}(\bm{\Theta}_{t,1})]\tran,\ldots,[\operatorname{vec}(\bm{\Theta}_{t,\bar{G}})]\tran \bigr]\tran $. We should set $T=\bar{G}\bar{M}^2$ to guarantee the recovery of $\mathbf{\tilde{Z}}[l]$. Then, the lower bound for the MSE of the estimated channel can be attained with the optimal matrix $\bm{\Phi}$ derived as $\bm{\Phi}=\bm{\Psi}_1 \otimes \bm{\Psi}_2 $ with $\bm{\Psi}_1=\mathbf{F}_{\bar{G}}$ and 
$ \bigl[ \bm{\Psi}_2\bigr]_{:,(m-1)\bar{M}+n} = \operatorname{circshift}\Bigl(\operatorname{vec}(\mathbf{U}_1),(n-1)\bar{M}\Bigr) \odot \bigl(\bigl[\mathbf{U}_2\bigr]_{:,m}\otimes \mathbf{1}_{\bar{M}}\bigr), \forall m,n \in \bar{M},$
where $\mathbf{U}_1= \mathbf{F}_{\bar{M}}$ and $\mathbf{U}_2 = \frac{1}{\sqrt{\bar{M}}}\mathbf{F}_{\bar{M}}$, in which $\mathbf{F}_{\bar{M}}$ is the $\bar{M} \times \bar{M}$ DFT matrix \cite{li2023channel}. This LS-based algorithm can be derived for channel estimation for fully-connected RIS by substituting $\bar{M}=M$ and $\bar{G}=1$.

\subsection{The Proposed Tensor Based Channel Estimation Framework}


\subsubsection{Preliminaries on the Tucker2 Decomposition \cite{TensorReview,Tucker2_V2X}}

\begin{figure}[t]  
	\centering
	\includegraphics[width=0.5\textwidth]{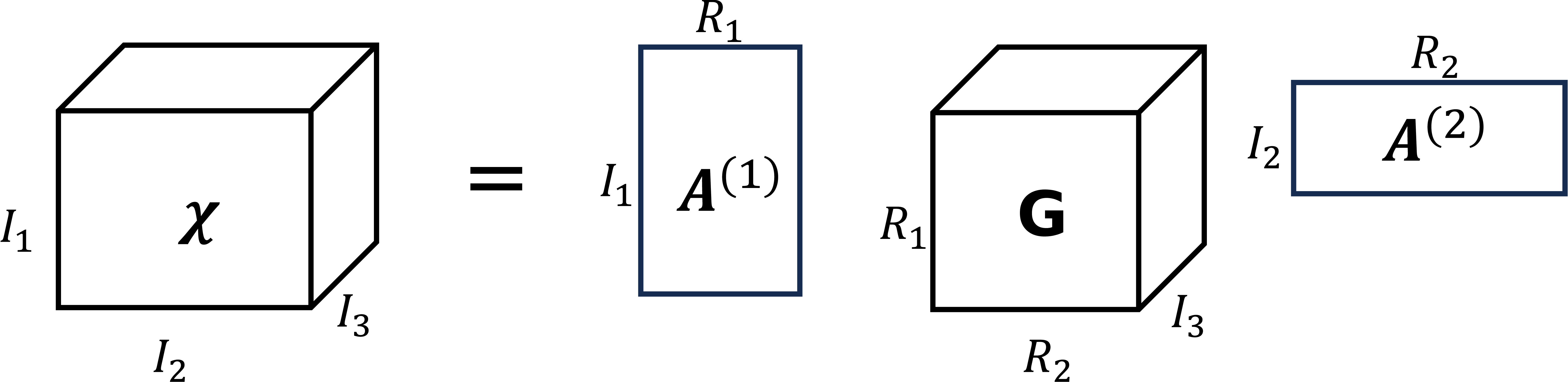}
	\caption{Representation of Tucker2 decomposition.}
	\label{fig:tucker2}
 \end{figure}

Due to the interconnected nature of BD-RIS, the composite channel matrix has many interdependent elements, which leads to a high computational complexity when estimating it using a conventional LS-based method. We propose to reduce the complexity by decomposing the composite channel using the Tucker2 decomposition method  \cite{TensorReview,Tucker2_V2X}. Tucker2 decomposition decomposes a multi-dimensional matrix into a set of matrices and one smaller core multi-dimensional matrix, as illustrated in Fig.~\ref{fig:tucker2}. It is used to model higher-order data by means of relatively small numbers of components for each of the two modes while linking them through a smaller core array of the same order as the original data. The model parameters are estimated in such a way that, given fixed numbers of components, the modelled data optimally resemble the actual data in the least squares sense. 

The Tucker2 decomposition breaks down the original third-order tensor $\tensor{X} \in \mathbb{C}^{I_1 \times I_2 \times I_3}$ into a third-order core tensor $\tensor{G} \in \mathbb{C}^{R_1 \times R_2 \times I_3}$ and two factor matrices, $\mathbf{A}^{(1)} \in \mathbb{C}^{I_1 \times R_1}$ and $ \mathbf{A}^{(2)} \in \mathbb{C}^{I_2 \times R_2}$, whose dimensions correspond to the indices of $\tensor{X}$, i.e., $I_1, I_2$, and $I_3$, and the multilinear ranks of $\tensor{X}$, i.e., $R_1$ and $R_2$, in the first and second modes, respectively. The Tucker2 decomposition of $\tensor{X}$ can be written using the $n$-mode product notation as 
  \begin{equation}\tensor{X} = \tensor{G} {\times}_1 \mathbf{A}^{(1)} {\times}_2 \mathbf{A}^{(2)} ,\label{T2}\end{equation}
  The mode-$1$ and mode-$2$ unfolding matrices of \eqref{T2}, according to the Kronecker product, in terms of the factor matrices and core tensor is given by
\begin{equation}[\tensor{X}]_{(1)} = \mathbf{A}^{(1)} \,[\tensor{G}]_{(1)} \left(\mathbf{I}_{I_3} \otimes \mathbf{A}^{(2)} \right)\tran \label{1},\end{equation} 
\begin{equation}[\tensor{X}]_{(2)} = \mathbf{A}^{(2)} \,[\tensor{G}]_{(2)} \left(\mathbf{I}_{I_3} \otimes \mathbf{A}^{(1)} \right)\tran \label{2}.\end{equation}

\subsubsection{Tensor Based Channel Estimation: Fully-connected RIS}
\label{Tucker2-BALS_FC}
We define a third-order tensor $\tensor{\tilde{Y}} \in \mathbb{C}^{N \times K \times T}$, which follows a Tucker2 decomposition and its $t$-th frontal slice can be interpreted as $\mathbf{\tilde{G}}_t[l]$ which is the noisy version of the ${\mathbf{G}_t[l]}$, UE-RIS-BS cascade channel in which derived after removing pilots symbols from receive signal $ {\mathbf{Y}}_t[l]$ as described in \eqref{noisycas}, and  it can be expressed as 
\begin{equation}
    \tensor{\tilde{Y}} = \tensor{Y} + \tensor{\tilde{V}},\label{noisyY}
\end{equation}
where $\tensor{Y}$ is the noiseless version of $\tensor{\tilde{Y}}$ and the $t$-th frontal slice of  $\tensor{\tilde{V}}$ can be interpreted as $\mathbf{\tilde{V}}_t[l]$.
The noiseless version,  $\tensor{Y}$ can be written using the n-mode product notation, as in \eqref{T2}, as follows: 
\begin{equation}\tensor{Y} = \tensor{W} {\times}_1 \mathbf{H} {\times}_2 \mathbf{E}[l]\tran ,\label{T2_real}\end{equation}
where $\bm{\Theta}_t(t=1,\ldots, T)$ is the $t$-th frontal slice of the core tensor $\tensor{W} \in \mathbb{C}^{M \times M \times T}$. We can write the following correspondences by comparing \eqref{T2} and \eqref{T2_real} as,
\begin{equation}\begin{array}{l}\left(\tensor{X},\tensor{G}, \mathbf{A}^{(1)}, \mathbf{A}^{(2)} \right) \leftrightarrow \left(\tensor{Y},\tensor{W}, \mathbf{H}, \mathbf{E}[l]\tran \right) \\ (I_1 ,I_2 ,I_3 ,R_1 ,R_2) \leftrightarrow (N,K,T,M,M). \end{array} \notag \end{equation}

 The mode-$1$ and mode-$2$ unfolding matrices of $\tensor{Y}$, is denoted by $[\tensor{Y}]_{(1)} \in \mathbb{C}^{N,KT}$ and$[\tensor{Y}]_{(2)} \in \mathbb{C}^{K,NT}$ with comparatively to \eqref{1} and \eqref{2}, respectively, in which satisfy the factorizations in terms of the factor matrices and core tensor as follows:
\begin{align}[\tensor{Y}]_{(1)} = \mathbf{H} [\tensor{W}]_{(1)} \left(\mathbf{I}_T \otimes \mathbf{E}[l]\tran \right)\tran,\label{h1}\\
[\tensor{Y}]_{(2)} = \mathbf{E}[l]\tran [\tensor{W}]_{(2)} \left(\mathbf{I}_T \otimes \mathbf{H} \right)\tran. \label{e1}\end{align}

The estimations of the channel matrices $\mathbf{H}$ and $\mathbf{E}[l]$ in the $l$-th coherence interval are obtained with the post-processed received signal $ \tensor{\tilde{Y}}$ given by \eqref{noisyY}, by fitting the third-order tensor $\tensor{\tilde{Y}} \in \mathbb{C}^{N \times K \times T}$ at BS to a Tucker2 model in an alternating way. This is achieved by optimizing the following two cost functions alternatively.
\begin{equation}\begin{array}{l}\mathbf{\hat{H}} =  \mathop {\text{argmin}}\limits_{\mathbf{H}}\left\| [\tensor{\tilde{Y}}]_{(1)} - \mathbf{H} [\tensor{W}]_{(1)} \left(\mathbf{I}_T \otimes \mathbf{E}[l]\tran \right)\tran\right\|_{\text{F}}^2 , \\ \mathbf{\hat{E}}[l]\tran =  \mathop {\text{argmin}}\limits_{\mathbf{E}[l]\tran} \left\| [\tensor{\tilde{Y}}]_{(2)} - \mathbf{E}[l]\tran [\tensor{W}]_{(2)} \left(\mathbf{I}_T \otimes \mathbf{H} \right)\tran \right\|_{\text{F}}^2 ,\end{array}\end{equation}
where $[\tensor{\tilde{Y}}]_{(1)}\in \mathbb{C}^{N,KT}$ is the 1-mode of $\tensor{\tilde{Y}}$ as in \eqref{h1} and  $[\tensor{\tilde{Y}}]_{(2)} \in \mathbb{C}^{K,NT}$ is the 2-mode of $\tensor{\tilde{Y}}$ as in \eqref{e1}.
Therefore, the estimations of the channel matrices $\mathbf{H}$ and $\mathbf{E}[l]$ in the $l$-th coherence interval are respectively given by
\begin{equation}\mathbf{\hat{H}} = [\tensor{\tilde{Y}}]_{(1)} \left[[\tensor{W}]_{(1)} \left(\mathbf{I}_T \otimes \mathbf{E}[l]\tran \right)\tran \right]^\dagger \label{h2} ,\end{equation}
\begin{equation}\mathbf{\hat{E}}[l]\tran = [\tensor{\tilde{Y}}]_{(2)} \left[[\tensor{W}]_{(2)} \left(\mathbf{I}_T \otimes \mathbf{H} \right)\tran \right]^\dagger \label{e2}.  \end{equation}

Each iteration of the BALS receiver goes through only two updating steps, for the estimation of $\mathbf{H}$ and $\mathbf{E}[l]$ according to equations \eqref{h2} and \eqref{e2}, respectively since $\tensor{W}$ is known.  {Here,
we select random values for $\bm{\Theta}$. Furthermore, each column of $\bm{\Theta}$ should be uncorrelated. }At each step, the estimation error is alternatively minimized concerning one channel matrix by keeping the other channel matrix fixed using its value in the previous step. This procedure is repeated iteratively, until convergence or the maximum number of algorithmic iterations, i.e., $i_{max}=30$\footnote{{By increasing the number of algorithmic iterations, training overhead can be reduced, ultimately leading to a sufficiently high estimation accuracy. Therefore, we need to select the maximum number of algorithmic iterations in such a way that it results in a sufficiently high estimation accuracy while ensuring the training overhead remains low.}}, is reached. The convergence is declared at the $i$-th iteration if the normalized mean square error (NMSE) of the residual error is less than the threshold $\kappa=1\times 10^{-6} $, i.e.,  $|e_{(i)}-e_{(i-1)}| \leq \kappa$ where the residual error, $e_{(i)}$ at the $i$-th iteration is defined as
\begin{equation}e_{(i)} = \left\| \tensor{\tilde{Y}} - \hat{\tensor{Y}}^{(i)} \right\|_{\text{F}}^2,\end{equation}
where $\hat{\tensor{Y}}^{(i)} $ denotes the reconstructed version of the $\tensor{\tilde{Y}}$, which is calculated using the estimated channel matrices in \eqref{h2} and \eqref{e2} after convergence. The Algorithm \ref{alg:1} summarizes the iterative BALS receiver for joint estimation of the channel matrices  $\mathbf{H}$ and $\mathbf{E}[l]$ in the $l$-th coherence interval.

\begin{algorithm}[t]
\caption{Iterative BALS Estimator: Fully-connected RIS}\label{alg:1}
\begin{algorithmic}[1]
\REQUIRE A feasible $\bm{\Theta}$, $\kappa$ and maximum number of iterations $i_{max}$ 
\STATE \textbf{Initialization:} Set $i=1$. Initialize randomly the the factor matrix $\mathbf{E}[l]_{(i=1)}$
\FOR{$i=1$ to $i_{max}$}
\STATE According to \eqref{h2}, obtain LS estimate for $\mathbf{H}$ \label{a2}
\STATE According to \eqref{e2}, obtain LS estimate for $\mathbf{E}[l]$ \label{a3}
\STATE \textbf{Until} $|e_{(i)}-e_{(i-1)}| \leq \kappa$  or $i \geq i_{max}$
 \ENDFOR
\STATE Remove the scaling ambiguities of $\mathbf{H}$  and $\mathbf{E}[l]$
 \ENSURE $\mathbf{\hat{H}}$ and $\mathbf{\hat{E}}[l]$
\end{algorithmic}
\end{algorithm}

\paragraph{Uniqueness and Identifiability Issues} 
\label{UIF}
{Generally, PARAFAC model gives unique decompositions. However,} the Tucker models are not essentially unique since unknown non-singular matrices can act as loading matrices\cite{TensorReview, TensorbasedCERelay}. It can shown by the equation \eqref{uniqueness} which is obtained using the property of the mode-n production.
\begin{align}
   \tensor{Y} &= \tensor{W} \times_{1} \mathbf{B}_1^{-1} \times_2 \mathbf{B}_2^{-1} \times_1 \mathbf{H}\mathbf{B}_1 \times_2 \mathbf{E}[l]\tran \mathbf{B}_2 \notag \\ &= \tensor{W} \times_1 \mathbf{H}\mathbf{B}_1\mathbf{B}_1^{-1} \times_2 \mathbf{E}[l]\tran \mathbf{B}_2 \mathbf{B}_2^{-1} \notag \\ &=  \tensor{W} \times_1 \mathbf{H}\times_2 \mathbf{E}[l]\tran,\label{uniqueness}
\end{align}
where $\mathbf{B}_1 \in \mathbb{C}^{M \times M}$ and $\mathbf{B}_2 \in \mathbb{C}^{M \times M}$ are the unknown non-singular matrices. To recover $\mathbf{H}$ and $\mathbf{E}[l]$ we need to eliminate this rotational freedom. It can be achieved if the $\mathbf{B}_1$ and $\mathbf{B}_2$ are identity matrices with complementary scale factors, i.e. $\mathbf{B}_1 = \frac{1}{\beta}\mathbf{I}_M$ and  $\mathbf{B}_2 =   \beta\mathbf{I}_M$. This means the recovery of $\mathbf{H}$ and $\mathbf{E}[l]$ up to a scalar factor\cite{ TensorbasedCERelay}.  To achieve unique decomposition we assume the $\beta$ is the first element of $\mathbf{H}$ and $\beta$ is known. We can estimate the $[\mathbf{H}]_{1,1}$ by only switching on the first element in RIS and the first antenna in BS before Tucker2-BALS estimation. 

The identifiability of Tucker2 is also crucial to recover $\mathbf{H}$ and $\mathbf{E}[l]$. In order to achieve identifiability it is needed to  $\left[[\tensor{W}]_{(1)} \left(\mathbf{I}_T \otimes \mathbf{E}[l]\tran \right)\tran\right]$ and $\left[[\tensor{W}]_{(2)} \left(\mathbf{I}_T \otimes \mathbf{H} \right)\tran \right]$ in equations \eqref{h2}, \eqref{e2} be full row rank. That means it needs to be satisfied $ M\leq min(NT, KT). $

\subsubsection{Tensor Based Channel Estimation: Group-connected RIS}
\label{Tucker2-BALS_GC}
During the pilot-based training phase for group-connected RIS, we divide the {training blocks} used for channel estimation into sub-groups such that {training blocks} for one group are $T_g = {T}/{\bar{G}}$. The minimum pilot length required for channel estimation increases compared to fully-connected RIS for the same number of elements at the RIS due to the separate channel estimation procedure for each group. In each sub-group, we only switch on one subgroup of RIS elements and switch off other RIS elements. A similar method as \ref{Tucker2-BALS_FC} is applied for each group independently for channel estimation in group-connected RIS. 

Therefore the estimations of the channel matrices $\mathbf{H}_g$ and $\mathbf{E}_g[l]$ in $g$-th sub-group in the $l$-th coherence interval are respectively given by
\begin{equation}\mathbf{\hat{H}}_g = [\tensor{\tilde{Y}}]_{(1)}^g \left[[\tensor{W}]_{(1)}^g \left(\mathbf{I}_{T_g} \otimes \mathbf{E}_g[l]\tran \right)\tran \right]^\dagger \label{h2G} ,\end{equation}
\begin{equation}\mathbf{\hat{E}}_g[l]\tran = [\tensor{\tilde{Y}}]_{(2)}^g \left[[\tensor{W}]_{(2)}^g \left(\mathbf{I}_{T_g}\otimes \mathbf{H}_g\right)\tran \right]^\dagger \label{e2G}. \end{equation}

Each iteration of the BALS receiver goes through only two updating steps, for the estimation of $\mathbf{H}_g$ and $\mathbf{E}_g[l]$ according to equations \eqref{h2G} and \eqref{e2G}, respectively since $\tensor{W}_g$ is known. The convergence criteria for the BALS algorithm for group-connected RIS is similar to the convergence criteria for fully-connected RIS which is mentioned in Section \ref{Tucker2-BALS_FC}. The Algorithm \ref{alg:2} summarizes the iterative BALS receiver for joint estimation of the channel matrices  $\mathbf{H}$ and $\mathbf{E}[l]$ in the $l$-th coherence interval for group-connected RIS.

\begin{algorithm}[t]
\caption{Iterative BALS Estimator: Group-connected RIS}\label{alg:2}
\begin{algorithmic}[1]
\REQUIRE Feasible $\bm{\Theta}_g ,g=1,\ldots,\bar{G}$, $\kappa$ and maximum number of iterations $i_{max}$ 
\FOR{$g=1$ to $\bar{G}$}
\STATE \textbf{Initialization:} Set $i=1$. Initialize randomly the the factor matrix $\mathbf{E}_g[l]_{(i=1)}$
\FOR{$i=1$ to $i_{max}$}
\STATE According to \eqref{h2G}, obtain LS estimate for $\mathbf{H}_g$ \label{a2G}
\STATE According to \eqref{e2G}, obtain LS estimate for $\mathbf{E}_g[l]$ \label{a3G}
\STATE \textbf{Until} $|e_{(i)}-e_{(i-1)}| \leq \kappa$  or $i \geq i_{max}$
 \ENDFOR
\STATE Remove the scaling ambiguities of $\mathbf{H}_g$  and $\mathbf{E}_g[l]$
 \ENDFOR 
 \ENSURE $\mathbf{\hat{H}}$ and $\mathbf{\hat{E}}[l]$
\end{algorithmic}
\end{algorithm}

 \paragraph{Uniqueness and Identifiability issues} 
The uniqueness and Identifiability issues Tucker2-BALS for the fully-connected RIS, which we discussed in Section \ref{UIF} are valid for group-connected RIS too. When considering uniqueness, the recovery of $\mathbf{H}_g$ and $\mathbf{E}_g[l]$ up to a scalar factor\cite{ TensorbasedCERelay}.  To achieve unique decomposition we assume the $\beta_g$ is the first element of $\mathbf{H}_g$ and $\beta_g$ is known. We can estimate the $[\mathbf{H}_g]_{1,1}$ by only switching on the first element in the $g$-th group in the RIS and the first antenna in BS before Tucker2-BALS estimation. when considering identifiability, it is needed to  $\left[[\tensor{W}]_{(1)}^g \left(\mathbf{I}_{T,g }\otimes \mathbf{E}_g[l]\tran \right)\tran \right]$ and $\left[[\tensor{W}]_{(2)}^g \left(\mathbf{I}_{T,g} \otimes \mathbf{H}_g \right)\tran \right]$ in equations \eqref{h2G}, \eqref{e2G} be full row rank. That means it needs to be satisfied $\bar{M} \leq min( NT_g, KT_g).$

\section{Channel Prediction for BD-RIS}
\label{channel prediction}
In this Section, we focus on the prediction phase in Fig.~\ref{fig:Transmission Block}.
\subsection{Channel Aging Model}
The channel aging property primarily arises due to user mobility, and this characteristic can be approximately described using the second-order statistics of the channel, i.e., through the autocorrelation function (ACF)\cite{YuanChannelAginML}. Considering the Rayleigh fading assumption, the discrete-time ACF for the fading channel coefficients is given by
\begin{equation}R[l]=J_{0}(2\pi f_{n}\vert l\vert),\end{equation} where $J_{0}(\cdot)$ is the zeroth-order Bessel function of the first kind, in which $\vert l\vert$ is the delay in terms of the number of coherence intervals, $f_n$ is the normalized Doppler frequency defined as  $f_{n}=f_{\mathrm{d}}T_s$.

The AR stochastic model can be used to model channel aging phenomena, requiring only the channel correlation matching property \cite{YuanChannelAginML}. Specifically, we can model time-varying UE-RIS channels of the considered BD-RIS-assisted MIMO system as a small-scale correlated fading series as follows:
\begin{equation} \mathbf{E}[l]=-\sum_{q=1}^{Q}a_{q}\mathbf{E}[l-q]+\boldsymbol{\omega}[l], \label{eq1} \end{equation}
where $\boldsymbol{\omega}[l]$ is an uncorrelated complex white Gaussian noise vector with zero mean and variance
\begin{equation} {\sigma^{2}_{\omega}}=R[0]+\sum_{q=1}^{Q}a_{q}R[-q], 
\label{eq2}\end{equation}
and the parameter $Q$ represents the order of the AR model.
The AR coefficients $\{a_{q}\}_{q=1}^{Q}$ are evaluated via the Levinson-Durbin recursion as
\begin{equation}
    {\mathbf a} = - {\mathbf R}^{-1}{\mathbf w},
    \label{eq3}
\end{equation}
where 
\begin{equation}
 {\mathbf R} =\left[\begin{matrix}
R[0] & R[- 1] & \cdots & R[- Q + 1]\cr R[1] & R[0] & \cdots & R[- Q + 2]\cr\vdots & \vdots & \ddots & \vdots\cr R[Q - 1] & R[Q - 2] & \cdots & R[0] \end{matrix}\right],
\label{eq4}
\end{equation}
\begin{equation}
{\mathbf a} = [\begin{matrix} 
a_{1} & a_{2} & \cdots & a_{Q}\end{matrix}]\tran,
\label{eq5}
\end{equation}
\begin{equation}
 {\mathbf w} = \left[\begin{matrix} R[1] & R[2] & \cdots & R[Q]\end{matrix}\right]\tran.
 \label{eq6}
\end{equation}

 The accuracy of the AR model improves with higher-order $Q$. The upper bound for $Q$ is equivalent to the number of coherence intervals used to collect CSI samples. 

Generating stable AR filters of high orders to accurately represent bandlimited channels can be challenging. A practical heuristic approach to address numerical issues was proposed in  \cite{AR_model}. The authors presented a strategy to enhance the conditioning of the autocorrelation matrix ${\bf R}$ by slightly increasing the values in its main diagonal with a small positive amount $\epsilon$. By selecting a suitable $\epsilon$, the lower bound of $\sigma^2_{\omega_k}$ can be significantly increased, enabling the stable computation of larger-order AR models. Here we assume $\epsilon = 0.1$. Therefore, this strategy is adopted in this paper. Following \cite{AR_model}, the first $Q+1$  ACFs of the resulting AR process can be reliably computed as  
\begin{equation}
\widehat{R}[q] = \begin{cases}
                   R[0] + \epsilon, & q = 0 \\
                   R[q], & q = 1,2,\ldots,Q.
                  \end{cases}
\end{equation}

\subsection{ML-Based Channel Prediction: CNN-AR Model} 
\label{CNN-ARFull}
\begin{figure}[t]  	
	\includegraphics[trim={0 0 14cm 0},width=0.7\textwidth]{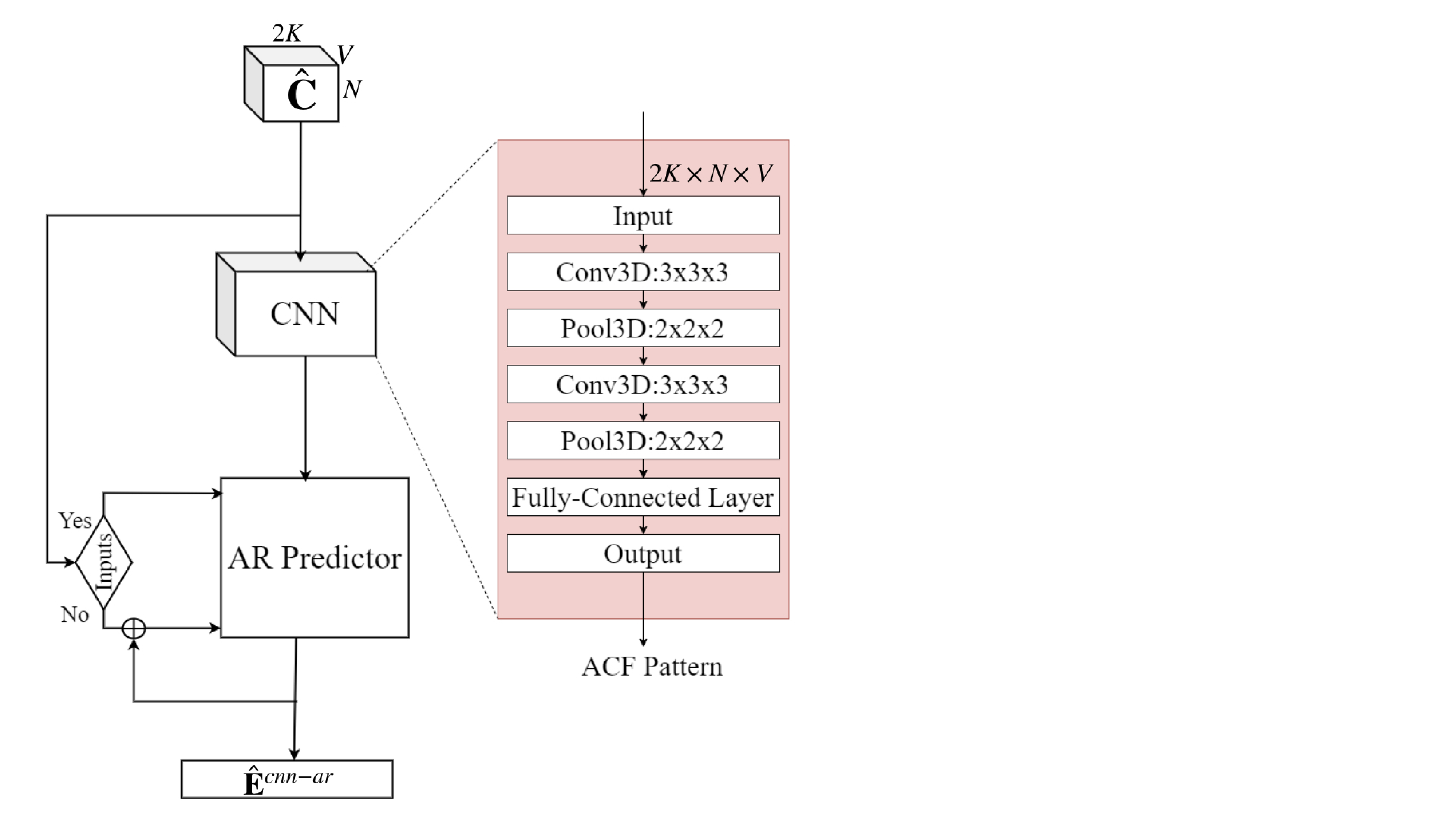}
	\caption{Architecture of the proposed CNN-AR predictor.}
	\label{fig:cnn-ar}
\end{figure} 

The performance of AR-based predictors generally improves with the increase of $Q$. The implication of this is that a simple AR predictor requires a large amount of CSI for high accuracy, which can be difficult to obtain in practice. To overcome these potential challenges, we propose a CNN-based prediction strategy to identify the channel aging pattern and determine accurate AR coefficients with arbitrary orders for a BD-RIS-assisted MIMO system. 

 To this end, a CNN model is used to extract the ACF pattern by treating the CSI data as image data. The diagram of the proposed CNN-AR predictor is shown in Fig.~\ref{fig:cnn-ar}. Our proposed CNN model can identify the ACF pattern in UE-RIS reflected channel when initial estimations in the pilot-based training phase are found by Tucker2-based channel estimation.

The estimated channels $\mathbf{\hat{E}}$, for $l=1,\cdots, V$, obtained during the pilot-based training phase are used as inputs for the implemented CNN model. Generally, $V$ should be equal to the order of the AR model, i.e.,$V=Q$. This training data is first preprocessed, in which the estimated complex channel vectors collected during $V$ coherence intervals are transformed into a real-valued matrix, as follows:
\begin{align}
    \mathbf{\hat{C}} = \begin{bmatrix}
         \mathrm{Re}\{\mathbf{\hat{E}}[1]\} & \cdots & \mathrm{Re}\{\mathbf{\hat{E}}[V]\} \\
    \mathrm{Im}\{\mathbf{\hat{E}}[1]\} & \cdots & \mathrm{Im}\{\mathbf{\hat{E}}[V]\}
    \end{bmatrix}.
\end{align}

The proposed CNN model is used to identify the ACF pattern of the input CSI values. Subsequently, the proposed framework utilizes the AR coefficients associated with the aging pattern obtained by the CNN model. It then forecasts the CSI for the upcoming first coherence interval, as follows: 
\begin{equation} \mathbf{\hat{E}}^{\mathrm{cnn-ar}}\left [{ l }\right] = - \sum_{q = 1}^{Q} {a_{q}} \mathbf{\hat{E}}\left[{l - q}\right].\end{equation}

The result for the current coherence interval is used to predict the CSI for the next $P-1$ intervals as follows:  
\begin{align}&\hspace {-2pc}\mathbf{\hat{E}}^{{\mathrm{cnn-ar}}} \left [{ {l + l'} }\right] \notag \\=&- \sum_{q = l' + 1}^{Q} {a_{q}{\mathbf{\hat{E}}}\left [{ {l + l' - q} }\right]} \notag \\&-\, \sum _{q' = 1}^{l'} {a_{q'}}{\mathbf{\hat{E}}^{{\mathrm{cnn-ar}}}\left [{ {l + l' - q'} }\right]},\;\;\; l'\in P.\end{align} When there is input CSI from pilot-based training phase, those inputs are used for prediction ('Yes' arrow direction for AR predictor in Fig.~\ref{fig:cnn-ar}), and when there is no inputs, previously predicted CSI are used for channel prediction ('No' arrow direction for AR predictor in Fig.~\ref{fig:cnn-ar}).

As for the model architecture, we utilize the tanh activation function in the convolutional layers and sigmoid in fully connected layers. We implement two hidden layers in full connection. Furthermore, we adopt the adaptive moment estimation (Adam) optimizer and employ the mean-squared error (MSE) loss function. 

\begin{figure}[t]  
	\centering	\includegraphics[width=0.5\textwidth]{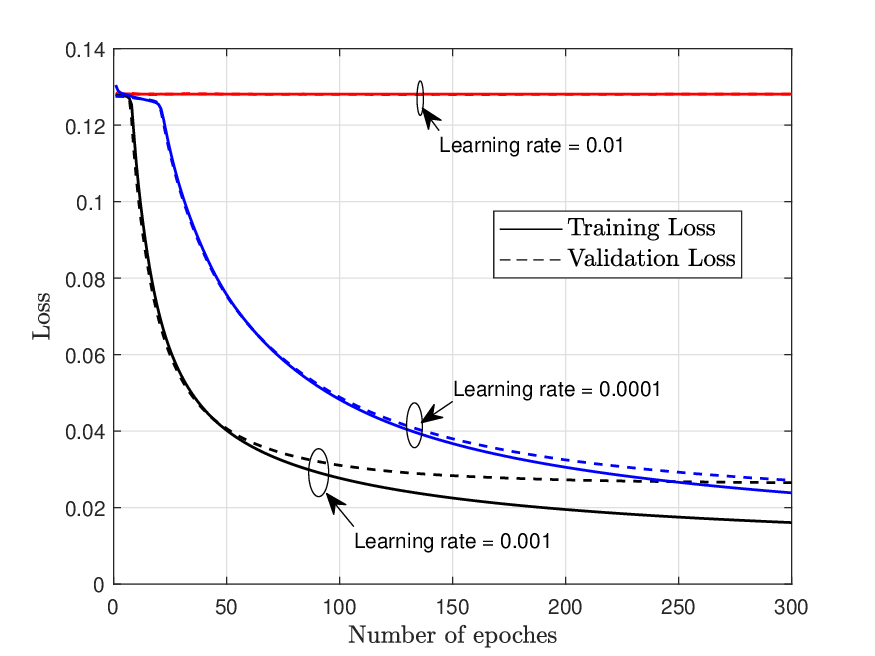}
	\caption{Training and validation loss of the CNN model in the CNN-AR predictor ($M=16$, fully-connected RIS).}
	\label{fig:convergence}
\end{figure}

Fig. \ref{fig:convergence} brings the training and validation losses of the CNN model in the CNN-AR predictor for different learning rates, considering a fully-connected RIS with $M=16$ reflecting elements. Similar curves can also be obtained for the group-connected RIS case, which is {omitted }due to space constraints. Fig. \ref{fig:convergence} shows that the rate $0.01$ is excessively high since the losses do not improve with the number of iterations. On the other extreme, the rate of $0.0001$ leads to a slower convergence. Therefore, we fix the learning rate to $0.001$ since it achieves the best performance. 

Training, validation, and testing datasets have been generated with $7\times 10^4$, $2\times 10^4$, and $1\times 10^4$ samples, respectively, for the CNN model in the channel prediction phase. Even though we use $10$ different $f_n$ values to train the CNN model, Our proposed CNN model is capable of adding more $f_n$ values in training. The ML model undergoes $300$ epochs while the early-stopping is employed to avoid overfitting during the training phase, with a batch size of $50$.

\section{RIS Phase Optimization and Performance Metrics}
\label{performance}
The proposed BALS-based channel estimation scheme with Tucker2 decomposition for BD-RIS-assisted systems can decompose the UE-RIS-BS channel into a product of lower dimensional matrices, thereby reducing the complexity of the channel estimation process. Moreover, the proposed channel prediction scheme integrates a CNN model with an AR predictor to take advantage of the channel correlation and channel aging phenomena to predict future channel instances without requiring pilot signals. Finally, we optimize the reflection matrix of the BD-RIS using the estimated/predicted CSI to analyze the performance of our proposed channel estimation scheme. We adopt four important metrics to analyze the performance of our proposed scheme, which are explained in the following.
\subsection{ {Average Downlink Sum Rate Performance} }
  {We use the downlink sum rate as the first performance analysis metric to examine the influence of the CSI which is estimated and predicted through our proposed schemes. we can use estimated uplink channels to obtain the downlink channel since we assume channel reciprocity. Therefore, we are interested in the optimization of the downlink weighted sum power to find the optimal reflection matrices that give the maximum downlink sum rate. }
 
  {We jointly optimize the precoding vector for the $k$-th UE, denoted as $\mathbf{u}_{k}$, and the passive reflection coefficients $\bm{\Theta}$ of the RIS, aiming at maximizing the weighted sum power at the $k$-th UE during the downlink data transmission phase in one coherence interval as follows:
 \begin{align}(\mathrm{P}1):&\max _{{\Theta} ,\:\mathbf {u}_k } \quad {\hspace{-2mm}\sum _{k^` = 1}^{K} {\alpha}_k P_d{\sum _{k = 1}^{K} {\left ({{\left |{(\mathbf{e}_k \bm{\Theta}\mathbf {H}\tran)\herm\mathbf {u}_k}\right|^{2}} }\right)} } }  \notag \\&\textrm {s.t.} \quad   \sum _{k = 1}^{K} \|\mathbf {u}_k\|^{2} = 1,\notag \\&\hphantom {\textrm {s.t.} \quad } \bm{\Theta} = \diag{(\bm{\Theta}_1,\cdots,\bm{\Theta}_G)}, \notag \\&\hphantom {\textrm {s.t.} \quad }   \bm {\Theta }_g\herm\bm {\Theta }_g = \mathbf{I}, \hspace{1mm} \bm {\Theta }_g = \bm {\Theta }_g\tran, \forall g, \label{opt1} \end{align}
where $P_d$ is the transmitted signal power and ${\alpha}_k > 0$ is power weight of the $k$-th UE. Here estimated CSI has been used to find the optimal reflection coefficients. We use a single-stream precoding vector to eliminate interference and find the optimal passive reflection coefficients using the closed-form solution proposed in \cite{closed_form_theta}.}

 {The downlink average downlink sum rate is defined as 
 \begin{equation} R = \mathbb{E} \left\{ { {\frac {{{T_{c}} - {P_{a}}}}{{{T_{c}}}}\sum _{k = 1}^{K} {{\log _{2}}\left ({{1 + {\mathrm {SIN}}{{\mathrm {R}}_{k}}} }\right)} } }\right\},\end{equation}
 where $\lambda=\frac {{{T_{c}} - {P_{a}}}}{{{T_{c}}}}$ is the data transmission coefficient which expresses the ratio of effective data transmission in the downlink in one coherence interval where $P_a$ is the average pilot overhead. Here  ${\mathrm {SIN}}{{\mathrm {R}}_{k}}$  is the signal-to-interference and noise ratio (SINR) for $k$-th UE which is defined as follows:
\begin{equation}
\mathrm{SINR}_k = \frac{{\alpha}_k P_d\left |{(\mathbf{e}_k \bm{\Theta}\mathbf {H}\tran)\herm\mathbf {u}_k}\right|^{2} }{{\alpha}_k P_d\sum _{i \neq 1}^{K} \left |{(\mathbf{e}_k \bm{\Theta}\mathbf {H}\tran)\herm\mathbf {u}_i}\right|^{2}+{\sigma}^2_k},
\end{equation}
where ${\sigma}^2_k$ is the noise variance at the $k$-th UE.}

\subsection{ Accuracy}
We use the NMSE to evaluate the channel estimation and prediction accuracy. The NMSE metric is computed as
 \begin{align} \mathrm{NMSE}\left [{l}\right]={\mathrm {E}}\left \{ 
 \frac{ \left \|{\mathbf{\hat{\Xi}} \left [{ l }\right] - {{\mathbf{{\Xi}}}}\left [{ l }\right]} \right \|_{2}^{2}}{ {{ {\left \|{ {{{\mathbf{{\Xi}}}}\left [{ l }\right]} }\right \|_{2}^{2}}}} } \right \},\end{align}
 where $\hat{\Xi}[l] = \mathbf{\hat{H}}\mathbf{\hat{E}}[l]$ is the estimated/predicted channel matrix in $l$-th coherence interval and ${\Xi}[l] = \mathbf{H}\mathbf{E}[l].$  
 

\subsection{Pilot Overhead Analysis}
\label{overhead}
When using conventional TDD channel estimation methods for the fully-connected RIS, the channel estimation of uplink channels demands an overall pilot length of $\bar{T}^{\text{FC-CONV}} = M^2(Q+P)$ for the total considered $Q+P$ coherence intervals, as demonstrated in \cite{li2023channel}, with $M$ denoting the number of RIS elements, $Q$ the number of coherence intervals in the pilot-based training phase, and $P$ the number of coherence periods in the prediction phase. On the other hand, our proposed channel estimation scheme for the for fully-connected RIS only demands a total pilot length of $\bar{T}^{\text{FC-PROP}}=(T+1)Q$ during $Q+P$ coherence intervals, where $T$ is the minimum pilot length required for the Tucker2-BALS algorithm. Now, considering the group-connected RIS, the channel estimation of uplink channels demands, in general, a pilot length of $\bar{T}^{\text{GC-CONV}} = (\bar{M}^2\bar{G})(Q+P)$ for the total considered $Q+P$ coherence intervals, if we are using conventional TDD channel estimation methods \cite{li2023channel}, recalling that $\bar{M}$ represents the number of RIS elements per group and $\bar{G}$ the number of groups. On the other hand, our proposed channel estimation scheme for the group-connected RIS only demands an overall pilot length of $\bar{T}^{\text{GC-PROP}}=(T+\bar{G})Q$ during the $Q+P$ coherence intervals. 

Note that there is no pilot consumption during the prediction phase in our proposed joint channel estimation and channel prediction algorithm, which should provide significant signalling savings. To compare the pilot overhead of the different channel estimation schemes in our numerical results, we consider the average of the overhead throughout the transmission block. The average pilot overhead for the total considered $Q+P$ coherence intervals is obtained as follows:
\begin{equation}
    P_a = \frac{\bar{T}^i}{Q+P},
\end{equation}
where $\bar{T}^i$ is the total pilot length for the given estimation strategy $i \in \{\text{FC-CONV, GC-CONV, FC-PROP, GC-PROP}\}$.

\subsection{Complexity Analysis}
\renewcommand{\arraystretch}{1.2}
\begin{table*}[t]
	\caption{Computational Complexity} 
	\centering 
	\begin{tabular}{|l|  l| l|} 
		\hline 
	    \textbf{BD-RIS architecture} &     \textbf{{Proposed (Tucker2-BALS+CNN-AR)}} &   \textbf{{Conventional (DFT-LS)}} \\
		\hline
             Fully-connected & $\mathcal{O}\bigl(iQ(M^2T(K+N)+2NKTM)+28514KMQ + 64(Q+1)+8192\bigr)$ & {$\mathcal{O}\bigl((Q+P)M^4(1+NK)\bigr)$}\\
             \hline
             $M=16$ & $7.5 \times 10^7$&{$6.1 \times 10^7$} \\
             \hline
             {$M=64$} & {$2.9\times 10^9$}&{$6.6 \times 10^9$}\\
             \hline
            Group-connected  & $\mathcal{O}\bigl(iQ\bar{G}(\bar{M}^2T(K+N)+2NKT\bar{M})+28514KMQ + 64(Q+1)+8192\bigr)$ & {$\mathcal{O}\bigl((Q+P)\bar{G}^2\bar{M}^4(1+NK)\bigr)$}\\
            \hline
            $\bar{M}=8, \bar{G}=2$ &$6.0 \times 10^7$&{$1.5 \times 10^7$} \\
            \hline
            {$\bar{M}=32, \bar{G}=2$} &{$9.6 \times 10^8$}&{$1.6 \times 10^9$} \\
            \hline
	\end{tabular}
	\label{table:complexity} 
\end{table*}
The per-iteration computational complexity of the Tucker2-BALS channel estimation algorithm mainly depends on calculating pseudo inverse and complex multiplication. 
\begin{enumerate}
     \item Fully-connected RIS: The per-iteration computational complexity of step 3 and step 4 of the Algorithm \ref{alg:1} is $\mathcal{O}(M^2KT+NKTM)$ and $\mathcal{O}(M^2NT+NKTM)$, respectively. Therefore the total computational complexity for Tucker2-BALS for fully-connected RIS is $\mathcal{O}\bigl(i(M^2T(K+N)+2NKTM)\bigr)$ where $i$ is the iterations needed for convergence of Tucker2-BALS.

    \item Group-connected RIS: The per-iteration, per group computational complexity of step 4 and step 5 of the Algorithm \ref{alg:2} is $\mathcal{O}(\bar{M}^2KT+NKT\bar{M})$ and $\mathcal{O}(\bar{M}^2NT+NKT\bar{M})$, respectively. Therefore the total computational complexity for Tucker2-BALS is $\mathcal{O}\bigl(i\bar{G}(\bar{M}^2T(K+N)+2NKT\bar{M})\bigr)$.
\end{enumerate}

The computational complexity of the CNN-AR channel prediction mainly depends on the calculation of the auto-correlation of the channel using the CNN model. The total computational complexity of the CNN model during the online testing process is $\mathcal{O}\bigl(28514KMQ + 64(Q+1)+8192\bigr)$. We ignore the computational complexity during the offline training process.

The total computational complexity for our proposed joint Tucker2-BALS-based channel estimation and CNN-AR-based channel prediction algorithm and the total computational complexity of conventional DFT-LS channel estimation scheme\cite{li2023channel} for both fully-connected RIS and group-connected RIS is presented in the Table \ref{table:complexity}. It is clearly visible that our proposed joint channel estimation and channel prediction algorithm archives low computational complexity {when the number of RIS elements increases}, compared to the existing channel estimation algorithms for BD-RIS for both fully-connected and group-connected architectures by analyzing the Table \ref{table:complexity}, {since the maximum power of $M$ is 2 in our proposed scheme while maximum power $M$ is 4 in the DFT-LS channel estimation scheme. A roughly ten-fold complexity reduction is observed for group-connected BD-RIS with the example case of $\bar{M}=32, \bar{G}=2$.}

\section{Numerical results}
\label{results}
\renewcommand{\arraystretch}{1.2}
\begin{table}[t]
	\caption{Simulation Parameters} 
	\centering 
	\begin{tabular}{|l| l| } 
		\hline 
	    Number of transmit antennas at BS, $N$  & $5$\\
		\hline
		Number of UEs, $K$ & $5$\\
	    \hline
           {Rician Factor} &  {2.8} \\
          \hline
        Carrier frequency, $f_c$ & $3$~GHz \\
        \hline 
        Number of time slots in one coherence interval, $T_c$  & $1\times 10^4$\\
        \hline
        Sampling duration, $T_s$ & $1\times 10^{-5}$~s \\
        \hline 
	\end{tabular}
	\label{table:simulation parameters} 
\end{table}
In this Section, we present insightful simulation results to assess the performance of our proposed joint channel estimation and channel prediction algorithm.  The adopted simulation parameters are listed in Table \ref{table:simulation parameters}.  

\subsection{{Average Downlink Sum Rate Performance Analysis}}
\begin{figure}[t]
    \centering
      \begin{subfigure}{1\columnwidth}
	\includegraphics[width=\textwidth]{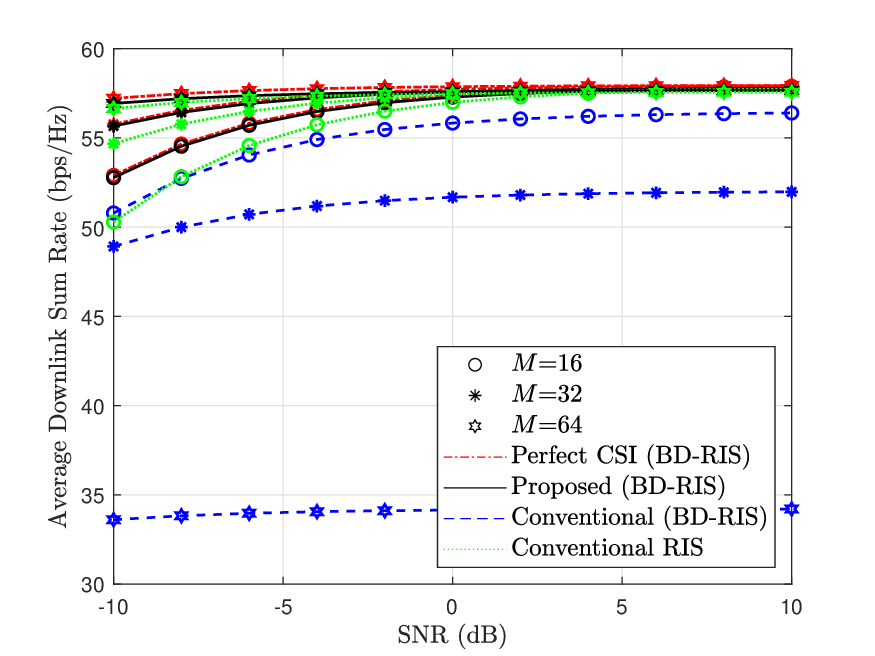}
         \subcaption{Fully-connected RIS}
         \label{fig:Rate_a}
	\end{subfigure}
        \begin{subfigure}{1\columnwidth}
		\includegraphics[width=\textwidth]{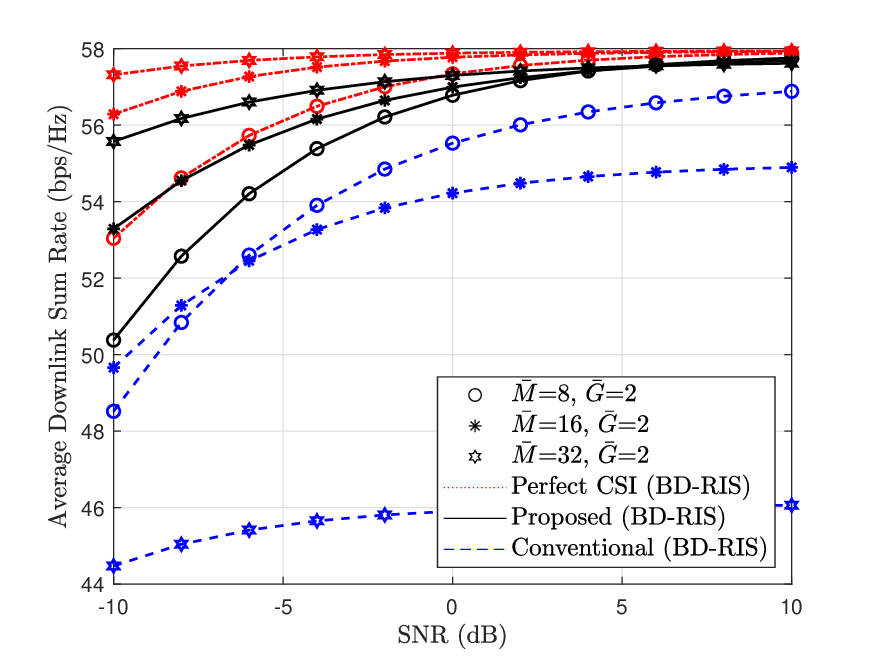}
  \subcaption{Group-connected RIS}
	\end{subfigure} 
 \caption{ {Comparison of average downlink sum rate with the estimated CSI using the proposed channel estimation scheme (Tucker2-BALS + CNN-AR) and using the conventional channel estimation scheme (DFT-LS) \cite{li2023channel} ($f_n =0.005$).}}
	\label{fig:Rate}
\end{figure}

We consider the average of {$R$} over the considered coherence intervals. Fig.~\ref{fig:Rate} shows the comparison of the average downlink sum rate with the estimated CSI using the proposed channel estimation scheme and using the conventional DFT-LS channel estimation scheme\cite{li2023channel} for fully-connected RIS and group-connected RIS. It shows that the sum rate is improved with the downlink signal-to-noise ratio (SNR). According to   Fig.~\ref{fig:Rate},  10-50\% average downlink sum rate gain can be achieved with the estimated CSI using the proposed channel estimation scheme when compared to existing channel methods and for BD-RIS in both fully-connected and group-connected architectures. When we increase the number of RIS elements, the average sum rate decays with the estimated CSI using conventional methods at high SNR due to the increase of $P_a$. However, the average sum rate with the estimated CSI using our proposed methods only increases when we increase $M$. Moreover, the average downlink sum rate which is achieved with the estimated CSI using the proposed channel estimation scheme is almost similar to the perfect CSI scenario for both architectures. {Furthermore, Fig. \ref{fig:Rate_a} shows that the average sum rate of conventional diagonal RIS  achieves a lower sum rate than the BD-RIS. We use PARAFAC decomposition for conventional RIS channel estimation which was proposed by the authors in \cite{PARAFAC_CE_RIS}.}

\subsection{ Accuracy}

\begin{figure}[t]
    \centering
      \begin{subfigure}{1\columnwidth}
	\includegraphics[width=\textwidth]{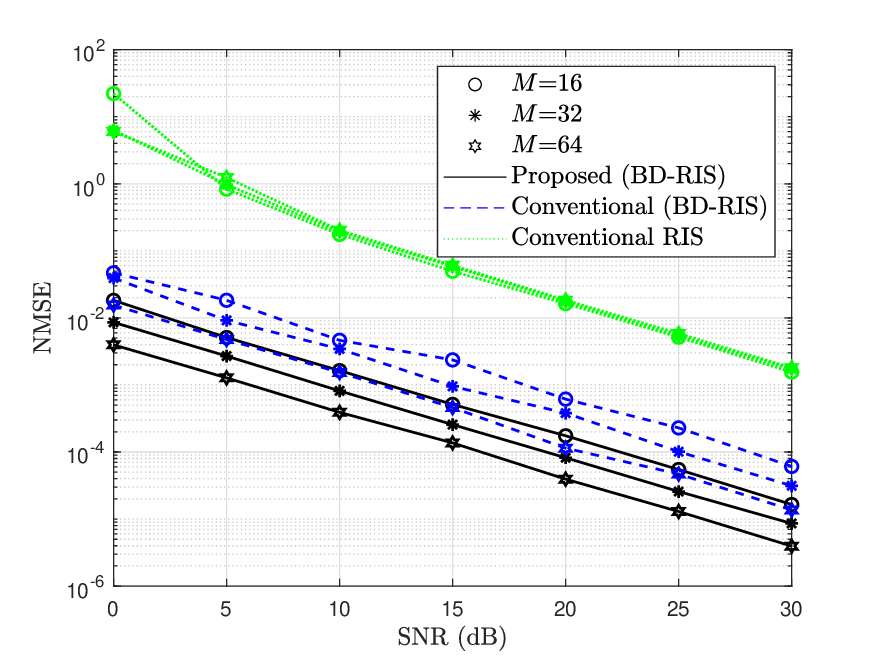}
         \subcaption{Fully-connected RIS}
         	\label{fig:NMSETuckervsDFT_a}
	\end{subfigure}
        \begin{subfigure}{1\columnwidth}
		\includegraphics[width=\textwidth]{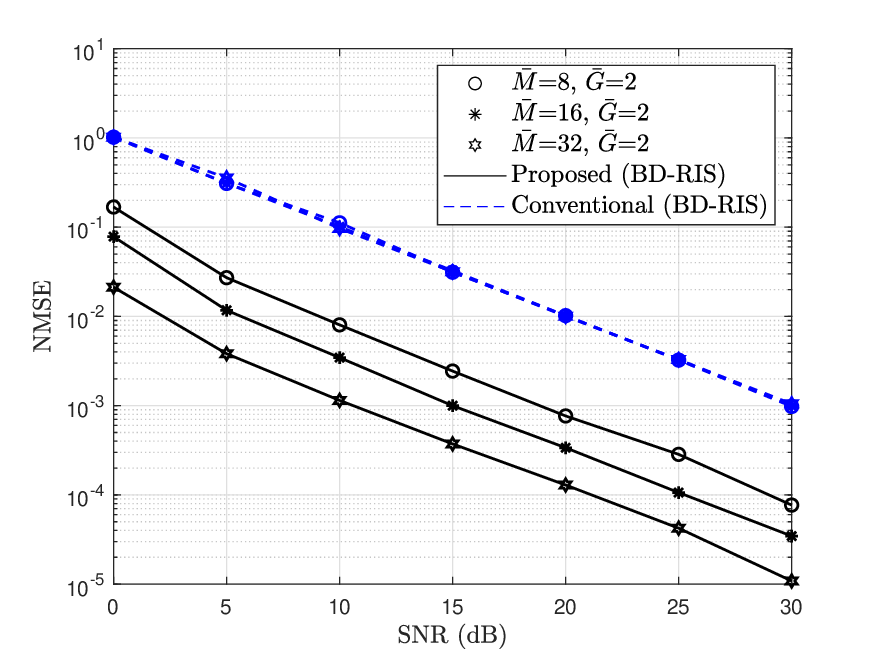}
  \subcaption{Group-connected RIS}
	\end{subfigure} 
  \caption{{Comparison of estimation accuracy of proposed channel estimation scheme (Tucker2-BALS) vs conventional channel estimation scheme (DFT-LS) \cite{li2023channel}.}}
	\label{fig:NMSETuckervsDFT}
\end{figure}
 
The NMSE is chosen to evaluate the estimation accuracy performance. Fig.~\ref{fig:NMSETuckervsDFT}  shows the comparison of estimation NMSE among the proposed channel estimation scheme and conventional channel estimation scheme \cite{li2023channel} for fully-connected RIS and group-connected RIS. The estimation accuracy increases with the SNR and the number of elements at the RIS.  According to  Fig.~\ref{fig:NMSETuckervsDFT} our proposed Tucker2-BALS achieves more than an order of magnitude lower NMSE with much fewer pilots (up to 56 times less) of pilots compared to existing methods.  {Furthermore, Fig. \ref{fig:NMSETuckervsDFT_a} shows the NMSE of PARAFAC decomposition-based channel estimation for conventional diagonal RIS \cite{PARAFAC_CE_RIS}. Estimation accuracy is reduced for conventional RIS compared to BD-RIS.}

\begin{figure}[t]  
    \centering
    \includegraphics[width=\columnwidth]{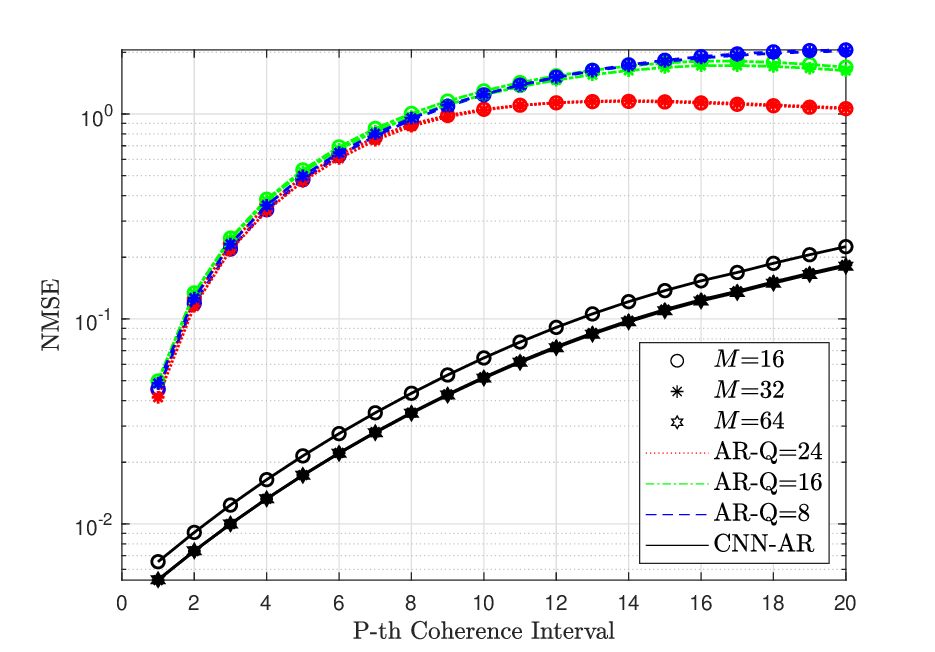}
    \caption{{Comparison of prediction NMSE among the AR predictors and CNN-AR for the first $P$ intervals for fully-connected RIS (Inputs for CNN-AR are found with the proposed Tucker2-BALS algorithm when SNR $= 20$~dB, $f_n =0.005$).} }
    \label{fig:NMSECNN_ARvsARFull}
\end{figure}

Fig.~\ref{fig:NMSECNN_ARvsARFull} shows the comparison of prediction NMSE among the AR predictors, and CNN-AR for the first $P$ intervals for fully-connected RIS. Similar behavior as Fig.~\ref{fig:NMSECNN_ARvsARFull} can be observed for the comparison of prediction NMSE among the AR predictors, and CNN-AR for the first $P$ intervals for group-connected RIS too.  According to Fig.~\ref{fig:NMSECNN_ARvsARFull}, the CNN-AR predictor has high prediction accuracy compared to AR estimators. Moreover, according to Fig.~\ref{fig:NMSECNN_ARvsARFull} comparison between $Q=8,16$ and $24$ shows that by expanding the order, the prediction accuracy is hardly improved in conventional AR predictor.  The reason behind the huge improvement in the prediction accuracy of the CNN-AR predictor is that the real-time calculation of the AR coefficient based on a limited number of CSI data inputs is not accurate enough. In contrast, the CNN-AR model can find the channel variation pattern and load the pre-computed AR coefficients in arbitrary order. 

\begin{figure}[t]
    \centering
    \includegraphics[width=\columnwidth]{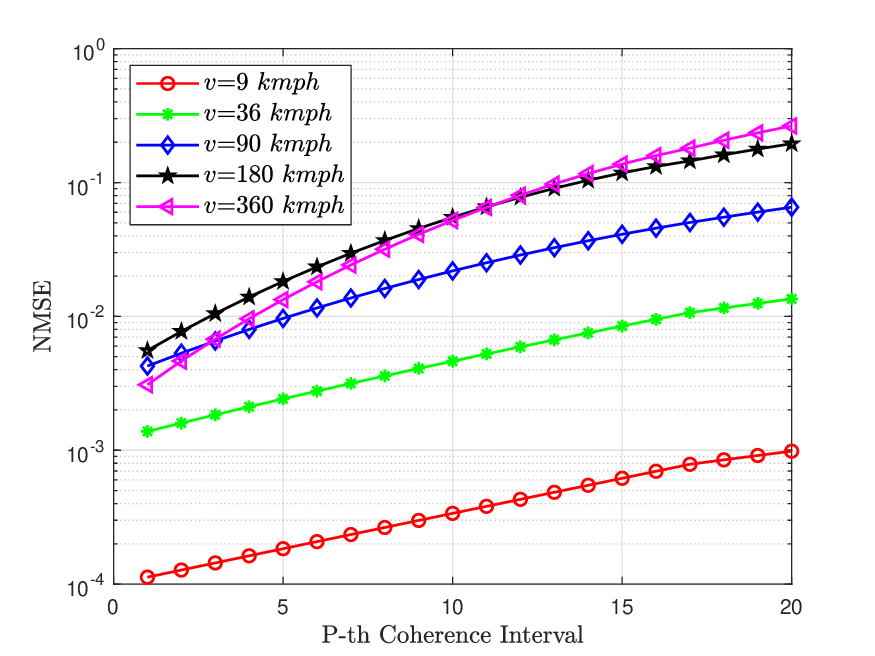}
    \caption{{Comparison of prediction NMSE of CNN-AR for different UE velocity ($v$) values when SNR $= 20$~dB, $M=16$, fully-connected RIS.}}
	\label{fig:fn}
\end{figure}
Fig.~\ref{fig:fn} shows the comparison of prediction NMSE of CNN-AR at the $P$-th coherence interval after the estimation phase for different UE velocity ($v$) values.  We considered five different $f_n$ values as shown in Fig.~\ref{fig:fn} which indicates the UE velocities from $9$~kmph to $360$~kmph  According to the Fig.~\ref{fig:fn}  our proposed CNN-AR channel predictor performs well in low UE velocities, i.e., for low channel aging values. The channel coherence time becomes shorter with higher UE velocity and hence the channel changes faster rendering the prediction less accurate. 
\subsection{Pilot Overhead Analysis}

\begin{figure}[t]
    \centering
     \begin{subfigure}{\columnwidth}
    \begin{subfigure}{0.5\columnwidth}
		\includegraphics[width=1\linewidth]{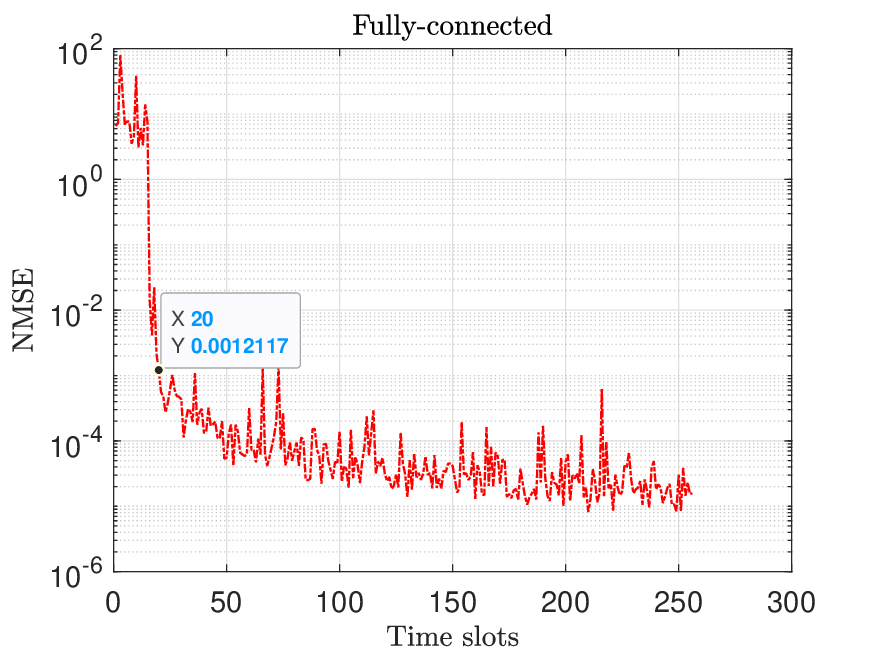}
    \end{subfigure}\hspace{-3mm}
    \begin{subfigure}{0.5\columnwidth}
  \includegraphics[width=1\linewidth]{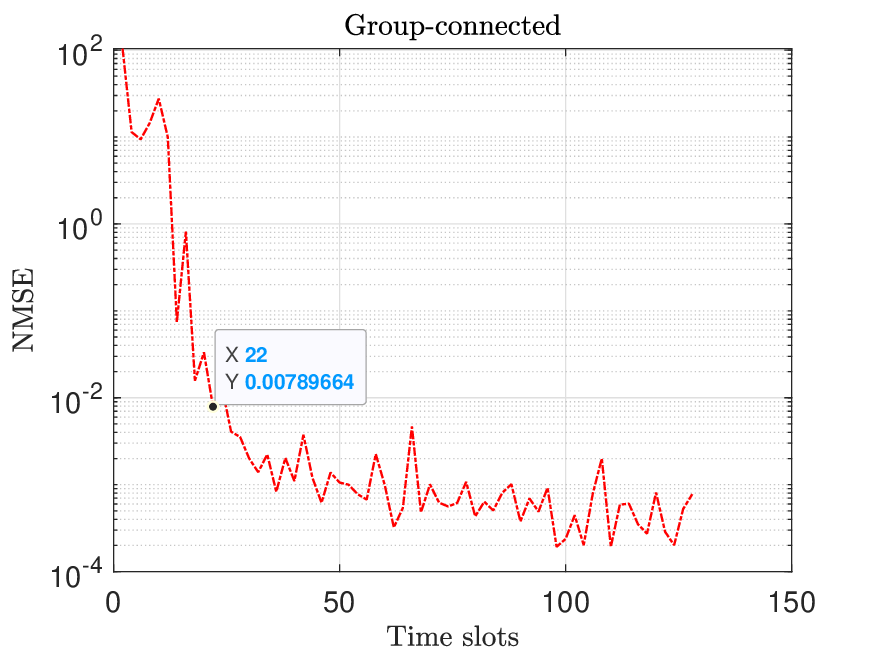}
    \end{subfigure}
    \subcaption{$M=16$}
    \end{subfigure}   
    \begin{subfigure}{\columnwidth}
    \begin{subfigure}{0.5\columnwidth}
	\includegraphics[width=1\linewidth]{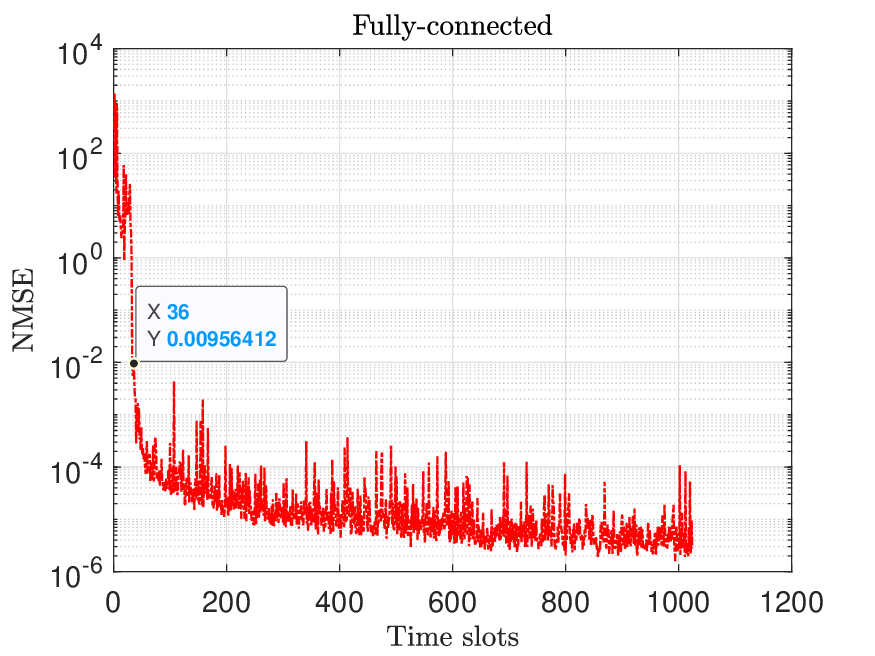}
    \end{subfigure}\hspace{-3mm}
    \begin{subfigure}{0.5\columnwidth}	\includegraphics[width=1\linewidth]{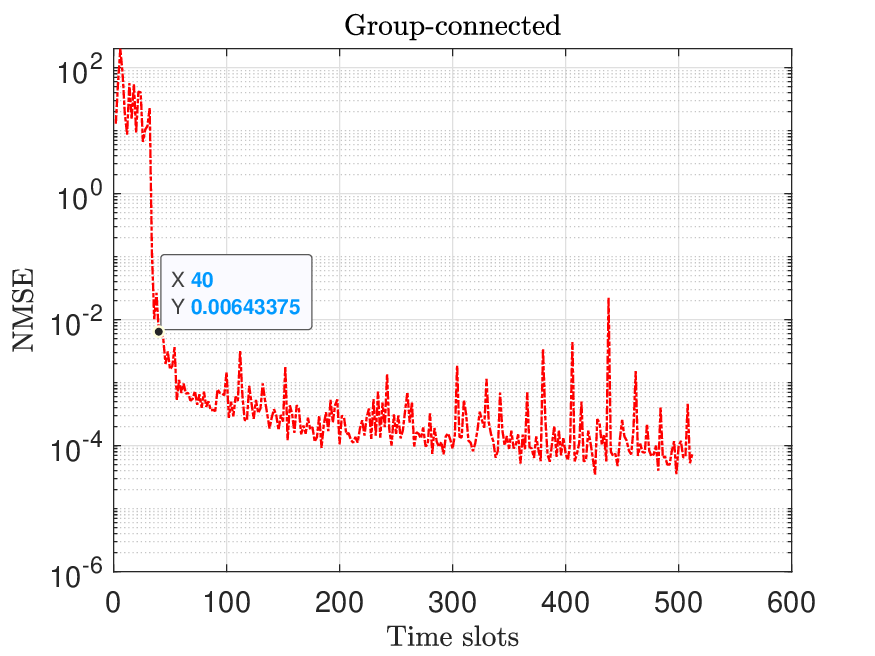}
    \end{subfigure}
    \subcaption{$M=32$}
    \end{subfigure}
    \begin{subfigure}{\columnwidth}
        \begin{subfigure}{0.5\columnwidth}
	\includegraphics[width=1\linewidth]{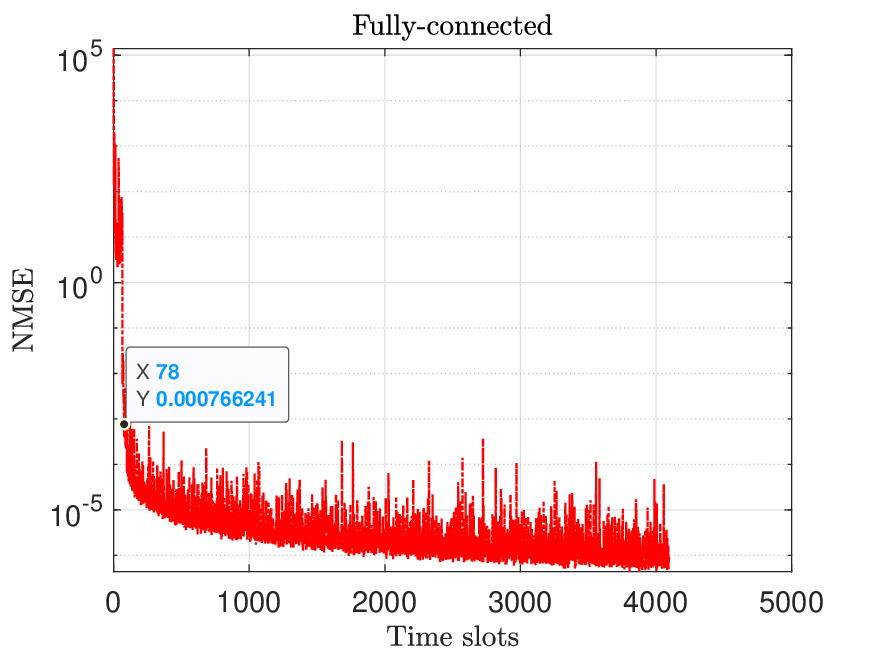}
    \end{subfigure}\hspace{-3mm}
    \begin{subfigure}{0.5\columnwidth}
		\includegraphics[width=1\linewidth]{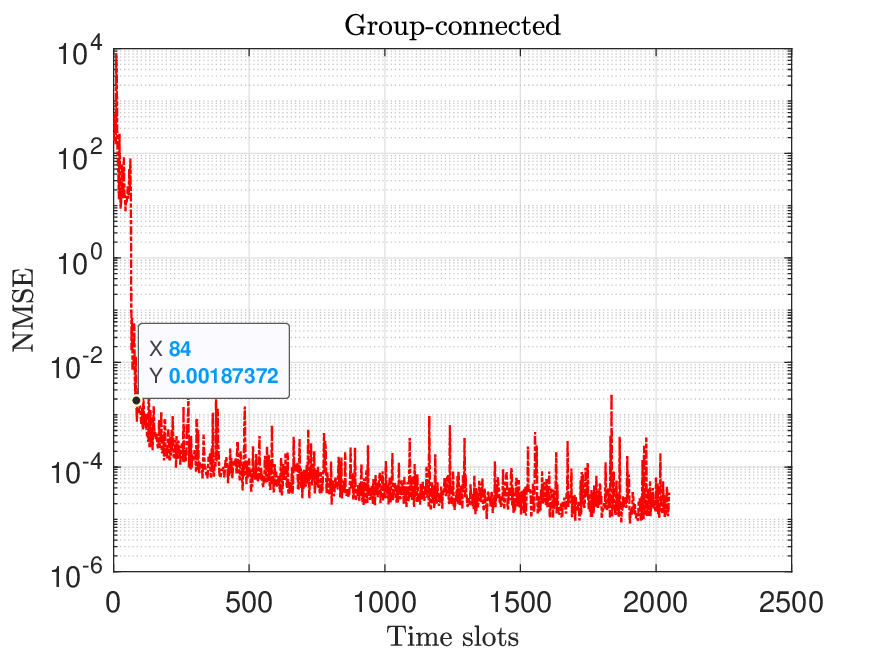}
    \end{subfigure}
    \subcaption{$M=64$}
    \end{subfigure}
\caption{{NMSE vs {training blocks} for Tucker2-BALS when SNR $= 20$~dB.} }
\label{fig:OptimalT}

\end{figure}

The impact of the pilot length on the estimation accuracy Tucker2-BALS during the pilot-based training phase is shown in Fig.~\ref{fig:OptimalT}.  It shows that with less number of {training blocks} it can be converge with low NMSE values with the Tucker2-BALS channel estimation for both fully-connected and group-connected RIS. Finding the optimum $T$ (the minimum pilot length required by the Tucker2-BALS) analytically is mathematically intractable, and hence we resort to numerical methods for the simulations. Therefore, with the help of Fig.~\ref{fig:OptimalT} we can find optimal values for $T$ when NMSE starts to converge. Table \ref{table:Pilot Overhead} shows the optimal values of $T$ for fully-connected and group-connected RIS we considered. Using those values we can calculate the average pilot overhead as mentioned in Section \ref{overhead} which is also presented in Table \ref{table:Pilot Overhead}. 94 - 98\% ({88}- 97\%) can be observed for the fully-connected (group-connected) BD-RIS architecture. The lower overhead reduction for the group-connected architecture is due to the fact that a separate channel estimation procedure is required for each group.

\renewcommand{\arraystretch}{1.5}
\begin{table*}[t]
	\caption{Comparison of the Average Pilot Overhead: Proposed vs. Conventional CSI Estimation Methods} 
	\centering 
	\begin{tabular}{|l| c| c| c| c| } 
		\hline 
	   \textbf{BD-RIS architecture} & \textbf{Minimum Pilot Length}  & $P_a$ \textbf{(Proposed)} & $P_a$ \textbf{(Conventional) }& \textbf{Pilot overhead reduction} \\
		\hline
             Fully-connected: $M=16$ & $T={20}$ & ${21}\frac{Q}{Q+P}$ &$256$ &{ 94.95\%} \\
             \hline
             Fully-connected: $M=32$ & $T={36}$ &  ${37}\frac{Q}{Q+P}$& $1024$ & {97.78\%}\\
             \hline
             Fully-connected: $M=64$ & $T={78}$ &  ${79}\frac{Q}{Q+P}$& $4096$ & {98.81\%}\\
		\hline
            Group-connected: $\bar{M}=8, \bar{G}=2$ & $T={22}$ &  ${24}\frac{Q}{Q+P}$ & $128$ & {88.46\%}\\
            \hline
            Group-connected: $\bar{M}=16, \bar{G}=2$ & $T={40}$ &  ${42}\frac{Q}{Q+P}$ &$512$ & {94.95\%}\\
            \hline
            Group-connected: $\bar{M}=32, \bar{G}=2$ & $T={84}$ &  ${86}\frac{Q}{Q+P}$ &$2048$ & {97.42\%}\\
            \hline
	\end{tabular}
	\label{table:Pilot Overhead} 
\end{table*}

\begin{figure}[t]  
    \centering
    \includegraphics[width=\columnwidth]{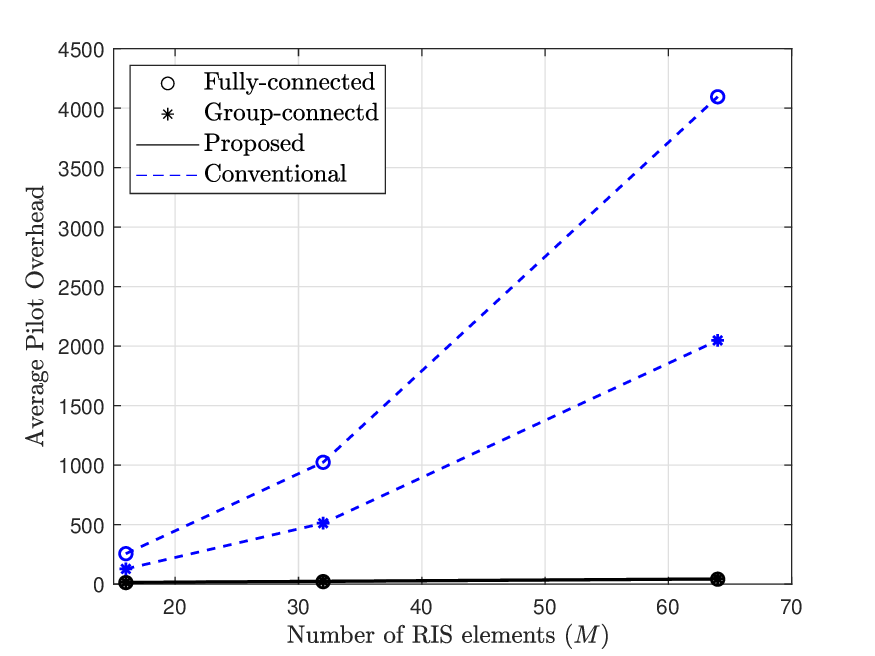}
    \caption{{Comparison of average pilot overhead when SNR $= 20$~dB. }}
    \label{fig:pilotOverhead}
\end{figure}

The comparison of the average pilot overhead for the proposed channel estimation scheme with the average pilot overhead for conventional DFT-LS channel estimation scheme for fully-connected RIS and group-connected RIS is shown in Fig.~\ref{fig:pilotOverhead}, which proves that our proposed scheme has low pilot overhead. In summary, our proposed joint channel estimation and channel prediction framework significantly reduces the channel estimation complexity and overhead; and demonstrates robustness to both types of BD-RIS architectures, i.e., fully-connected and group-connected.

\section{Conclusion}
\label{conclusion}
Accurate knowledge of CSI is essential in BD-RIS-assisted systems to optimize the RIS reflection matrix. Conventional channel estimation schemes have significant overhead and complexity due to the non-diagonal nature of the reflection matrix in BD-RIS. Complexity and overhead further increase with channel aging.  In this paper, we proposed a novel efficient joint Tucker2-BALS-based channel estimation and CNN-AR-based channel prediction solution to address the challenges associated with channel estimation in correlated multi-user, fully-connected, and group-connected BD-RIS-assisted MIMO systems with channel aging. The proposed Tucker2-BALS algorithm can estimate and decompose the channel with high accuracy. Also, the implemented CNN-AR approach successfully identified aging patterns and provided enhanced estimates of wireless channels, showcasing its high-precision estimation accuracy. Moreover, our extensive numerical analysis establishes the superiority of our approach in achieving a significantly low pilot overhead of up to $98\%$ pilot overhead reduction compared to existing methods. Furthermore, the results also show that our approach has low computational complexity and a high average downlink sum rate when compared to existing methods.

\bibliographystyle{IEEEtran}
\bibliography{IEEEabrv,main}{}

\end{document}